\documentclass[twocolumn]{aastex631}
\usepackage{amsmath}
\usepackage{multirow}

\graphicspath{{./figures/}}

\newcommand{\civ}{C\,{\sc iv}}
\newcommand{\heii}{He\,{\sc ii}}
\newcommand{\feii}{Fe\,{\sc ii}}
\newcommand{\oiii}{[O\,{\sc iii]}}

\newcommand{\mgii}{Mg\,{\sc ii}}
\newcommand{\ha}{H$\alpha$}
\newcommand{\hb}{H$\beta$}

\shorttitle{DESI SMBH MASS}
\shortauthors{Pan et al.}

\begin{document}
\title{Iron-corrected Single-epoch Black Hole Masses of DESI Quasars at low redshift}

\author[0000-0003-0230-6436]{Zhiwei Pan}
\affiliation{Department of Astronomy, School of Physics, Peking University, Beijing 100871, China}
\affiliation{Kavli Institute for Astronomy and Astrophysics, Peking University, Beijing 100871, China}

\author[0000-0003-4176-6486]{Linhua Jiang}
\affiliation{Department of Astronomy, School of Physics, Peking University, Beijing 100871, China}
\affiliation{Kavli Institute for Astronomy and Astrophysics, Peking University, Beijing 100871, China}

\author[0000-0001-9457-0589]{Wei-Jian Guo}
\affiliation{Key Laboratory of Optical Astronomy, National Astronomical Observatories, Chinese Academy of Sciences, Beijing 100012, China}

\author[0000-0002-1234-552X]{Shengxiu Sun}
\affiliation{Department of Astronomy, School of Physics, Peking University, Beijing 100871, China}
\affiliation{Kavli Institute for Astronomy and Astrophysics, Peking University, Beijing 100871, China}

\author[0000-0002-2949-2155]{Małgorzata Siudek}
\affiliation{Institute of Space Sciences, ICE-CSIC, Campus UAB, Carrer de Can Magrans s/n, 08913 Bellaterra, Barcelona, Spain}

\author{Jessica Nicole Aguilar}
\affiliation{Lawrence Berkeley National Laboratory, 1 Cyclotron Road, Berkeley, CA 94720, USA}

\author[0000-0001-6098-7247]{Steven Ahlen}
\affiliation{Physics Dept., Boston University, 590 Commonwealth Avenue, Boston, MA 02215, USA}

\author{David Brooks}
\affiliation{Department of Physics \& Astronomy, University College London, Gower Street, London, WC1E 6BT, UK}

\author{Todd Claybaugh}
\affiliation{Lawrence Berkeley National Laboratory, 1 Cyclotron Road, Berkeley, CA 94720, USA}

\author[0000-0002-1769-1640]{Axel  de la Macorra}
\affiliation{Instituto de F\'{\i}sica, Universidad Nacional Aut\'{o}noma de M\'{e}xico,  Cd. de M\'{e}xico  C.P. 04510,  M\'{e}xico}

\author{Peter Doel}
\affiliation{Department of Physics \& Astronomy, University College London, Gower Street, London, WC1E 6BT, UK}

\author{Enrique Gaztañaga}
\affiliation{Institut d'Estudis Espacials de Catalunya (IEEC), 08034 Barcelona, Spain}
\affiliation{Institute of Cosmology and Gravitation, University of Portsmouth, Dennis Sciama Building, Portsmouth, PO1 3FX, UK}
\affiliation{Institute of Space Sciences, ICE-CSIC, Campus UAB, Carrer de Can Magrans s/n, 08913 Bellaterra, Barcelona, Spain}

\author[0000-0003-3142-233X]{Satya Gontcho A Gontcho}
\affiliation{Lawrence Berkeley National Laboratory, 1 Cyclotron Road, Berkeley, CA 94720, USA}

\author{Stephanie Juneau}
\affiliation{NSF NOIRLab, 950 N. Cherry Ave., Tucson, AZ 85719, USA}

\author[0000-0003-3510-7134]{Theodore Kisner}
\affiliation{Lawrence Berkeley National Laboratory, 1 Cyclotron Road, Berkeley, CA 94720, USA}

\author{Andrew Lambert}
\affiliation{Lawrence Berkeley National Laboratory, 1 Cyclotron Road, Berkeley, CA 94720, USA}

\author[0000-0003-1838-8528]{Martin Landriau}
\affiliation{Lawrence Berkeley National Laboratory, 1 Cyclotron Road, Berkeley, CA 94720, USA}

\author[0000-0001-7178-8868]{Laurent Le Guillou}
\affiliation{Sorbonne Universit\'{e}, CNRS/IN2P3, Laboratoire de Physique Nucl\'{e}aire et de Hautes Energies (LPNHE), FR-75005 Paris, France}

\author[0000-0003-4962-8934]{Marc Manera}
\affiliation{Departament de F\'{i}sica, Serra H\'{u}nter, Universitat Aut\`{o}noma de Barcelona, 08193 Bellaterra (Barcelona), Spain}
\affiliation{Institut de F\'{i}sica d’Altes Energies (IFAE), The Barcelona Institute of Science and Technology, Campus UAB, 08193 Bellaterra Barcelona, Spain}

\author[0000-0002-4279-4182]{Paul Martini}
\affiliation{Center for Cosmology and AstroParticle Physics, The Ohio State University, 191 West Woodruff Avenue, Columbus, OH 43210, USA}
\affiliation{Department of Astronomy, The Ohio State University, 4055 McPherson Laboratory, 140 W 18th Avenue, Columbus, OH 43210, USA}
\affiliation{The Ohio State University, Columbus, 43210 OH, USA}

\author[0000-0002-1125-7384]{Aaron Meisner}
\affiliation{NSF NOIRLab, 950 N. Cherry Ave., Tucson, AZ 85719, USA}

\author{Ramon Miquel}
\affiliation{Instituci\'{o} Catalana de Recerca i Estudis Avan\c{c}ats, Passeig de Llu\'{\i}s Companys, 23, 08010 Barcelona, Spain}
\affiliation{Institut de F\'{i}sica d’Altes Energies (IFAE), The Barcelona Institute of Science and Technology, Campus UAB, 08193 Bellaterra Barcelona, Spain}

\author[0000-0002-2733-4559]{John Moustakas}
\affiliation{Department of Physics and Astronomy, Siena College, 515 Loudon Road, Loudonville, NY 12211, USA}

\author{Adam Myers}
\affiliation{Department of Physics \& Astronomy, University  of Wyoming, 1000 E. University, Dept.~3905, Laramie, WY 82071, USA}

\author{Claire Poppett}
\affiliation{Lawrence Berkeley National Laboratory, 1 Cyclotron Road, Berkeley, CA 94720, USA}
\affiliation{Space Sciences Laboratory, University of California, Berkeley, 7 Gauss Way, Berkeley, CA  94720, USA}
\affiliation{University of California, Berkeley, 110 Sproul Hall \#5800 Berkeley, CA 94720, USA}

\author[0000-0001-7145-8674]{Francisco Prada}
\affiliation{Instituto de Astrof\'{i}sica de Andaluc\'{i}a (CSIC), Glorieta de la Astronom\'{i}a, s/n, E-18008 Granada, Spain}

\author{Graziano Rossi}
\affiliation{Department of Physics and Astronomy, Sejong University, Seoul, 143-747, Korea}

\author[0000-0002-9646-8198]{Eusebio Sanchez}
\affiliation{CIEMAT, Avenida Complutense 40, E-28040 Madrid, Spain}

\author{Michael Schubnell}
\affiliation{Department of Physics, University of Michigan, Ann Arbor, MI 48109, USA}
\affiliation{University of Michigan, Ann Arbor, MI 48109, USA}

\author[0000-0002-6588-3508]{Hee-Jong Seo}
\affiliation{Department of Physics \& Astronomy, Ohio University, Athens, OH 45701, USA}

\author{David Sprayberry}
\affiliation{NSF NOIRLab, 950 N. Cherry Ave., Tucson, AZ 85719, USA}

\author[0000-0003-1704-0781]{Gregory Tarlé}
\affiliation{University of Michigan, Ann Arbor, MI 48109, USA}

\author{Benjamin Alan Weaver}
\affiliation{NSF NOIRLab, 950 N. Cherry Ave., Tucson, AZ 85719, USA}

\author[0000-0002-6684-3997]{Hu Zou}
\affiliation{National Astronomical Observatories, Chinese Academy of Sciences, A20 Datun Rd., Chaoyang District, Beijing, 100012, P.R. China}

\begin{abstract}

We present a study on the possible overestimation of single-epoch supermassive black hole (SMBH) masses in previous works, based on more than 55,000 type 1 quasars at $0.25 < z < 0.8$ from the Dark Energy Spectroscopic Instrument (DESI). We confirm that iron emission strength serves as a good tracer of the Eddington ratio, and estimate SMBH masses using an iron-corrected $R$--$L$ relation for \hb{}, where $R$ is the broad line region size and $L$ is the continuum luminosity. Compared to our measurements, previous canonical measurements without the iron correction are overestimated by a factor of 1.5 on average. The overestimation can be up to a factor of 5 for super-Eddington quasars. The fraction of super-Eddington quasars in our sample is about 5\%, significantly higher than 0.4\% derived from the canonical measurements. Using a sample featuring both \hb{} and \mgii{} emission lines, we calibrate \mgii-based SMBH masses using iron-corrected, \hb-based SMBH masses and establish an iron-corrected $R$--$L$ relation for \mgii. The new relation adds an extra term of $-0.34R_{\mathrm{Fe}}$ to the $R$--$L$ relation, where $R_{\mathrm{Fe}}$ denotes the relative iron strength. We use this formula to build a catalog of about 0.5 million DESI quasars at $0.6<z<1.6$. If these iron-corrected $R$-$L$ relations for \hb{} and \mgii{} are valid at high redshift, current mass measurements of luminous quasars at $z\ge6$ would have been overestimated by a factor of 2.3 on average, alleviating the tension between SMBH mass and growth history in the early universe.

\end{abstract}


\section{Introduction}  \label{sec:intro}
Quasars are generally accepted to be powered by the gravitational potential energy extracted from matter falling toward supermassive black holes (SMBHs). The study of quasar luminosity function and black hole mass function over cosmic time has enriched our understanding of physical processes related to the growth history of SMBHs \citep{Richards_2006,Kelly_2013,ShenXuejian_2020,PanZhiwei_2022,WuJin_2022,YangJinyi_2023,LiWenxiu_2023,FanXiaohui_2023}. Reliable SMBH mass measurements are of fundamental importance in determining the properties of quasars and SMBHs. Strong and broad emission lines originating from broad line regions (BLRs), including \ha{}, \hb{}, \mgii{}, and \civ{}, are frequently used to measure SMBH masses. Usually, we assume that the BLR gas moves virially under gravity \citep{Peterson_1999}, then the central SMBH mass can be estimated by the BLR size $R_{\mathrm{BLR}}$ and the velocity width $\Delta V$:
\begin{equation} \label{eq:M_RdeltaV}
M=f\frac{R_{\mathrm{BLR}}\Delta V^2}{G},
\end{equation}
where $f$ is the virial factor related to the geometry and kinematics of the BLR, and $G$ is the gravitational constant. Reverberation mapping (RM) measures $R_{\mathrm{BLR}}$ via a time lag between the continuum and line flux (see \citealt{Peterson_1993} and \citealt{Cackett_2021} for reviews). The emission-line profile of quasar spectra is used to measure $\Delta V$, which has been widely represented by the full width at half-maximum (FWHM) and the line dispersion $\sigma_{\mathrm{line}}$.

Although SMBH masses estimated from RM are usually robust, such a technique requires rich observational resources. Astronomers have tried to develop easier methods \citep[e.g.,][]{Vestergaard_2002} in the past few decades. For example, we can use single-epoch spectra to replace multi-epoch spectra, and thus use the empirical relation between the BLR size and the continuum luminosity (hereafter $R$--$L$ relation) to replace the time lag when estimating  $R_{\mathrm{BLR}}$. This type of scaling relations is not as robust as RM, so the RM results have been used to calibrate single-epoch scaling relations. Among these relations, the $R_{\mathrm{BLR}}($\hb$)$--$L_{5100}$ relation is the most reliable one, where $L_{5100}$ is the monochromatic luminosity ($\lambda L_{\lambda}$) at 5100 \r{A} , since the Balmer lines \ha{} and \hb{} in the rest-frame optical range are the basis for the majority of low-redshift RM programmes \citep[e.g.,][]{Kaspi_2000,Bentz_2006,Bentz_2013,Grier_2017,WangShu_2020,MaQinchun_2023,Cho_2023}. \ha{} has been served as a substitute when \hb{} is unavailable \citep[e.g.,][]{Greene_2005,Grier_2017,Cho_2023}. It is less popular than \hb{} because it presents more observational challenges and its applicability is limited to a narrower redshift range.

In the rest-frame UV wavelength range, researchers have extended this approach to consider other emission lines such as \mgii{} and \civ{} \citep{Vestergaard_2002,Warner_2004,Vestergaard_2009,Le_2020,ShenYue_2024}. The scaling relations based on \mgii{} and \civ{} play a crucial role in estimating the SMBH masses of higher-redshift ($z>1$) Active Galactic Nuclei (AGNs) because \hb{} moves out of the observed-frame optical range. Several RM studies have been conducted to investigate the $R$--$L$ relations using \mgii{} and \civ{} \citep[e.g.,][]{Metzroth_2006,Kaspi_2007,Lira_2018,Hoormann_2019,YuZhefu_2021,YuZhefu_2023,ShenYue_2024}. These relations are not as tight as those from \hb{} or \ha{}. Consequently, UV mass estimators require calibration against \hb{}-based measurements, including time lags (representing the BLR size) and mass estimates \citep[e.g.,][]{Woo_2018,Le_2020}. While the calibrated relations generally align, the scatter of the \civ{}-based relation is significantly larger than those of the \mgii{}-based and \hb{}-based relations \citep[e.g.,][]{Trakhtenbrot_2012,Coatman_2017}.

In the past several years, there is accumulating evidence suggesting an increasing scatter in the $R$--$L$ relation \citep[e.g.,][]{ShenYue_2024}. Astronomers have identified some reasons, including accretion states that can influence the canonical $R$--$L$ relation \citep{WangJianmin_2014_spin,WangJianmin_2014_slimdisk,McHardy_2018,GRAVITY_2024}. In particular, the Super-Eddington Accreting Massive Black Hole (SEAMBH) campaign has uncovered that many AGNs with strong \feii{} and narrow \hb{} emission lines, possibly indicating high accretion rates, deviate from the traditional $R$--$L$ relation \citep{DuPu_2015,DuPu_2018}. The study by \citet{DuPu_2019} claimed that the flux ratio between the \feii{} and \hb{} lines is strongly correlated with the deviation from the $R$--$L$ relation. Furthermore, \citet{Woo_2024} observed that most of their AGNs fall below the canonical relation, with the deviation potentially linked to the Eddington ratio. They identified a tentative correlation involving the relative \feii{} strength. Spectro-interferometric results by \citet{GRAVITY_2024} and \citet{Abuter_2024} also appear to support the hypothesis that the Eddington ratio is indeed a factor influencing the $R$--$L$ relation.

In terms of theory, many assumptions underlie the canonical $R$--$L$ relation, including the isotropic assumptions, proportional assumptions, point source assumptions \citep{DuPu_2019}. These conditions are tailored to the standard disk model with a normal accretion rate, suggesting that dependences of the $R$--$L$ relation extend to the accretion state and even other properties. Consequently, the extended $R$--$L$ relation is proposed to incorporate the \feii{} strength, aiming to provide less biased single-epoch estimations of the \hb{}-based SMBH mass and accretion rate \citep{DuPu_2019,YuLiming_2020}.

The necessity for an iron correction in the \mgii{}-based SMBH mass estimations remains uncertain. \citet{DongXiaobo_2011} observed a strong correlation between the Eddington ratio and the intensity ratios of both ultraviolet and optical \feii{} emission to \mgii{}. In a related context, \citet{Martinez_2020} suggested that a smaller scatter can be achieved through a linear combination of $L_{5100}$ and the strength of UV \feii{}. Conversely, some studies claimed that current \mgii{} samples consistently adhere to the canonical 2-parameter $R$--$L$ relation \citep{Khadka_2022,YuZhefu_2023}. Further validation is crucial with a need for a more extensive and well-defined RM-calibrated sample in the future.

In this paper, we adopt the iron-corrected $R$--$L$ relation proposed by \citet{DuPu_2019}. Then we utilize the corrected \hb{}-based SMBH mass as a reference to calibrate the \mgii{}-based SMBH mass. Despite the absence of new RM results, we leverage the high spectral resolution, high data quality, and large sample size of the Dark Energy Spectroscopic Instrument (DESI) to investigate DESI SMBH mass estimation with iron correction. We will not distinguish between AGNs and quasars in the following since we focus on the luminous sources. We use FWHM to represent the velocity width $\Delta V$. The layout of the paper is as follows. In Section \ref{sec:data_measure}, we outline the data from DESI, detailing fitting procedures and spectral measurements. Section \ref{sec:results} presents our quasar sample and corresponding analyses. In Section \ref{sec:discuss}, we assess the robustness of our conclusions, make comparisons between different SMBH mass estimators, and discuss the implications. We summarize the paper in Section \ref{sec:sum}. Throughout this paper, magnitudes are expressed in the AB system. We adopt a $\Lambda$-dominated flat cosmology with $H_0=70$ km s$^{-1}$ Mpc$^{-1}$, $\Omega_{m}=0.3$, and $\Omega_{\Lambda}=0.7$.

\section{Data and Measurements}{} \label{sec:data_measure}
This section offers a concise overview of the DESI imaging survey and DESI spectroscopic urvey. Furthermore, we illustrate the steps involved in creating the parent sample and performing spectral fitting for this sample.

\subsection{DESI}  \label{sec:DESI}
DESI, a Stage IV ground-based dark energy experiment, utilizes the 4 m Mayall telescope at Kitt Peak National Observatory to investigate the expansion history via baryon acoustic oscillations and cosmic structure growth through redshift-space distortions \citep{
,DESI_2016_I,DESI_2016_II,DESI_2022,DESI_2024_III,DESI_2024_IV,DESI_2024_VI}. Equipped with 5,000 fibers on the focal plane, DESI is conducting the largest multiobject spectral survey, efficiently observing approximately 40 million galaxies and quasars across 14,000 square degrees during its 5-year mission \citep{DESI_2016_I,DESI_2016_II,Silber_2023,Miller_2023}. It reaches a limiting magnitude of about 23 mag in the $r$ band, enabling the discovery of faint and high-redshift quasars \citep{Chaussidon_2023}. Furthermore, DESI has good spectral resolutions in its three optical channels (blue: 3600--5900 \AA, green: 5660--7220 \AA, and red: 7470--9800 \AA) with respective resolutions of approximately $R\sim2700$, $R\sim4200$, and $R\sim4600$ \citep{DESI_2022}. 

Before starting the Main Survey, DESI launched a 5-month Survey Validation (SV) campaign in December 2020. The SV campaign consisted of three phases intending to validate the selection of the targets, the development of the operations \citep{Schlafly_2023}, and the further optimization over an area equal to 1\% of the Main Survey \citep{DESI_2024_Validation}. The entire SV data is publicly released as the DESI Early Data Release \citep{DESI_2024_EDR}. Detailed information about the selection and validation for quasars, bright galaxies, emission-line galaxies, and luminous red galaxies from the SV data can be found in recent DESI publications \citep[e.g.,][]{Yeche_2020,Chaussidon_2023,Raichoor_2023,YangJinyi_2023,ZhouRongpu_2023,LanTingwen_2023}. Furthermore, comprehensive insights into the pipelines designed for DESI target selections and spectral reduction are presented in \citet{Myers_2023} and \citet{Guy_2023}, respectively. 

DESI employed a combination of three methods for quasar classification during SV: Redrock, a template-fitting classifier (Bailey et al. in preparation); \mgii{} Afterburner, a broad \mgii{} line finder \citep{Chaussidon_2023}; and QuasarNET, a deep convolutional neural network classifier \citep{Busca_2018,Farr_2020}. Subsequently, visual inspection was implemented to assess spectroscopic quality and redshift reliability, which reduced misclassified quasars and enhanced the performance of standard pipelines \citep{Alexander_2023}.

The spectra used in this work were obtained from the internal dataset called Iron, which is processed by the DESI pipeline. The Iron dataset that includes data from the SV and Year 1 phase will be published as part of the DESI Data Release 1 (DR1; DESI Collaboration et al. in preparation). To ensure that the spectral type classified by the pipeline is a quasar with a reliable redshift, we specifically selected those with SPECTYPE == ``qso'' and ZWARN == 0.

\subsection{DESI Legacy Imaging Survey} \label{sec:DESI_legacy}
The target selection of DESI relies on the imaging data from the DESI Legacy Surveys \citep{ZouHu_2017,Dey_2019}. The DESI Legacy Imaging Survey covers about 14,000 square degrees of the sky, with 9900 square degrees in the North Galactic Cap and 4400 square degrees in the South Galactic Cap regions. It was comprised of three distinct projects: the Dark Energy Camera Legacy Survey \citep[DECaLS;][]{Dey_2019}, the Beijing-Arizona Sky Survey \citep[BASS;][]{ZouHu_2017}, and the Mayall z-band Legacy Survey \citep[MzLS;][]{Dey_2019}. This imaging survey employed three bands and achieved approximate AB magnitudes of $g = 24.0$, $r = 23.4$, and $z = 22.5$ \citep{Dey_2019}. The photometric magnitudes from the Legacy Survey were utilized in this work.

\subsection{Parent Sample} \label{sec:parentsample}
The ``Iron'' dataset, spanning from December 14th, 2020, to June 13th, 2022, contains approximately 1.4 million quasar spectra. In this preliminary study, our primary focus is on high-quality, bright quasars, so we used the following selection criteria: $0.25<z<0.8$, $r<20.5$ mag, and $\mathrm{TSNR_{QSO}}>25$, where $\mathrm{TSNR_{QSO}}$ is the template signal-to-noise ratio (S/N) for quasar class \citep{Guy_2023}. The redshift cuts of 0.25 and 0.8 are constrained by the presence of \mgii{} and \hb{}, respectively. We selected bright quasars to minimize the influence from the host galaxy. As a result, we collected a total of 55,192 quasars as our parent sample. We then performed data preprocessing that includes correcting the spectra for galactic extinction and shifting the spectra to the rest frame. We adopted the galactic extinction curve proposed by \citet{Fitzpatrick_1999} with an assumed value of $R_{\mathrm{V}}$= 3.1.

\subsection{Spectral Fitting} \label{sec:spectral}
DASpec\footnote{\url{https://github.com/PuDu-Astro/DASpec}} is a versatile multi-component spectral fitting tool designed for AGNs, featuring a user-friendly graphical interface. It uses the popular Levenberg-Marquardt algorithm and the Simulated Annealing algorithm for the least squares fitting of the curves. Because this software offers the capability to tie or fix parameters during the fitting process, we utilized DASpec to perform individual spectral fitting within the optical and UV spectral regions for our parent sample. 

\begin{figure}
\epsscale{1.2}
\plotone{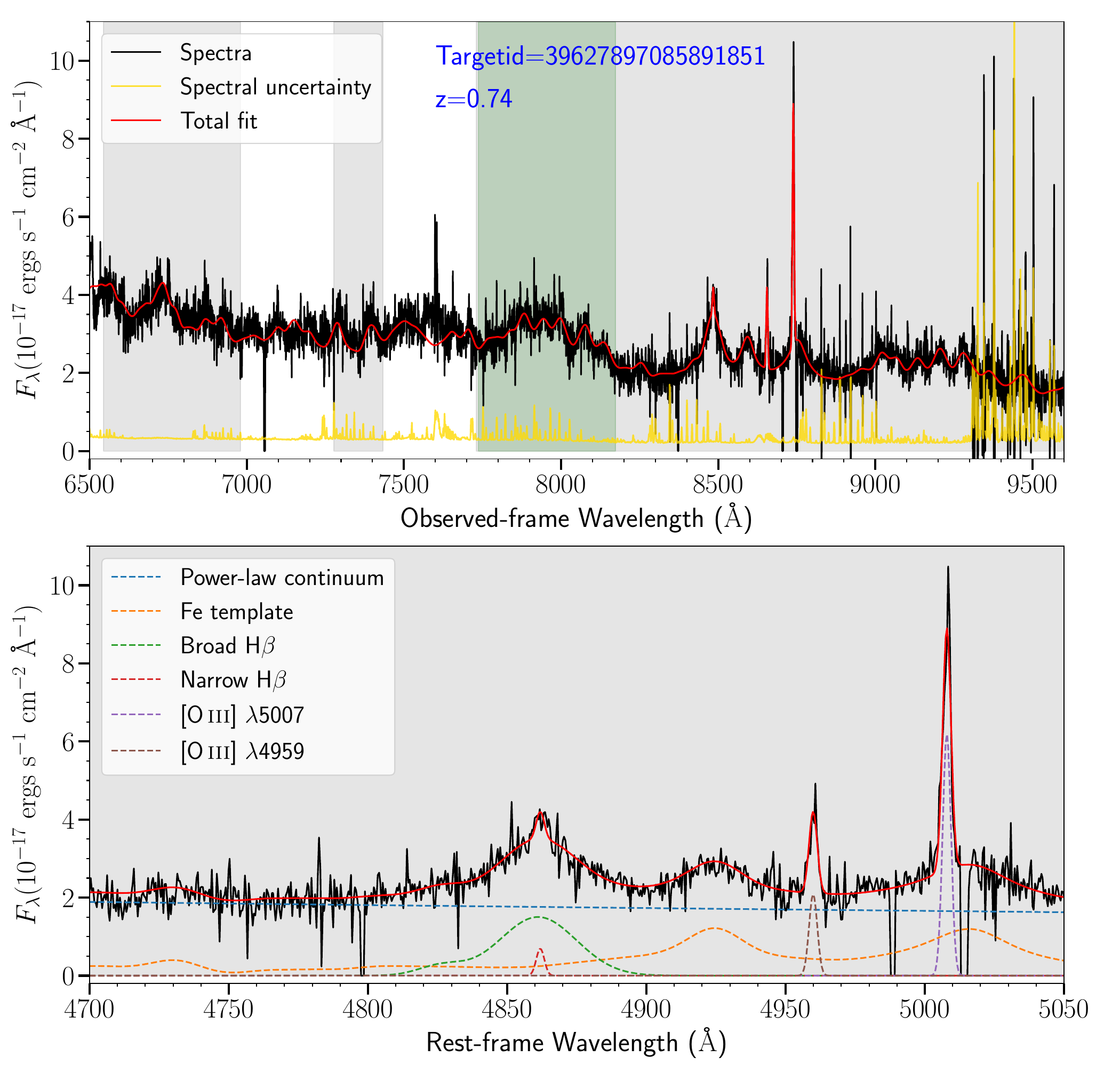}
\caption{A fitting with strong iron strength in the rest-frame optical region. In the upper panel, the black solid line represents the observed-frame spectrum following correction for galactic extinction, and the yellow solid line represents the spectral uncertainty. The red solid line illustrates the optimal total fit.  The gray shaded area illustrates the fitting windows and the green shaded area illustrates the wavelength range used for the iron flux calculation. In the zoom-in lower panel, the dashed lines denote the detailed components in the rest frame, including the power-law continuum, the \feii{} template, the broad \hb{} component, the narrow \hb{} component, \oiii\ $\lambda5007$, and \oiii\ $\lambda4959$. In this case, the \heii{} line is too weak to be displayed, so it is not shown here.} \label{fig:fit_example_opt}
\end{figure}

In the optical spectral region, our analysis comprises the following components: (1) a power-law continuum, (2) an \feii{} pseudo-continuum template developed by \citet{Boroson_1992}, (3) two Gaussian profiles to characterize the broad \hb{} component, (4) individual Gaussian profiles for narrow emission lines, such as \oiii\ $\lambda\lambda4959,5007$ and the narrow \hb{} component, and (5) a single Gaussian profile to model the broad \heii{} line. Notably, all narrow-line components for each object share the identical velocity width and shift. Additionally, \oiii\ $\lambda4959$ is constrained to have one-third of the \oiii\ $\lambda5007$ flux, following the prescription from \citet{Osterbrock_2006}. The fitting procedure is executed within specific wavelength windows in the rest frame, spanning 3750--4000 \AA, 4170--4260 \AA, and 4430--5500 \AA. The example of multicomponent spectral fitting in the optical region is illustrated in Figure \ref{fig:fit_example_opt}.

\begin{figure}
\epsscale{1.2}
\plotone{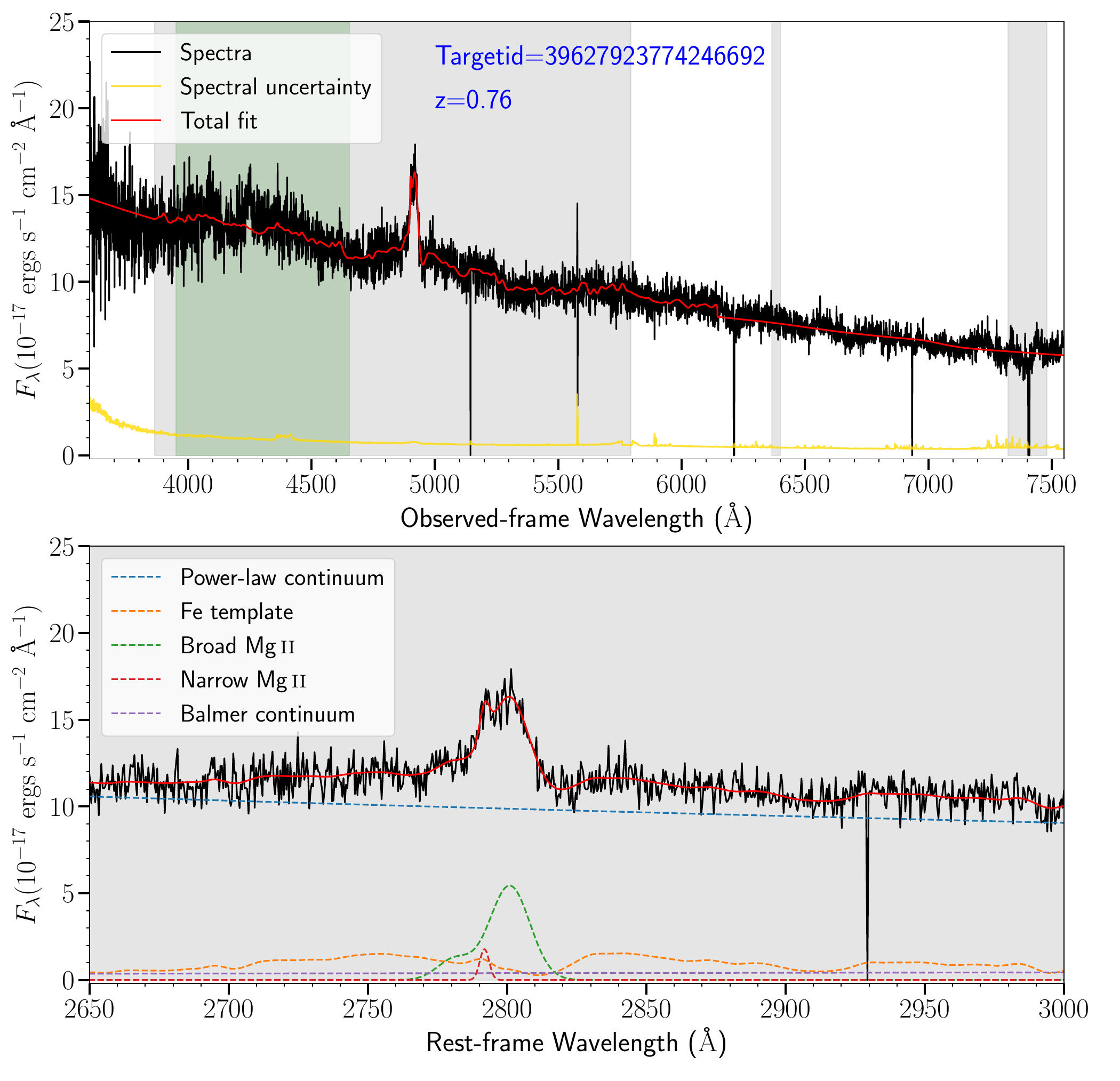}
\caption{A fitting with strong iron strength in the rest-frame UV region. In the upper panel, the black solid line represents the rest-frame spectrum following correction for galactic extinction, and the yellow solid line represents the spectral uncertainty. The red solid line illustrates the optimal total fit. The gray shaded area illustrates the fitting windows and the green shaded area illustrates the wavelength range used for the iron flux calculation. In the zoom-in lower panel, the dashed lines denote the detailed components in the rest frame, including the power-law continuum, the \feii{} template, the broad \mgii{} component, the narrow \mgii{} component, and the Balmer continuum.}\label{fig:fit_example_UV}
\end{figure}

In the UV spectral region, our analysis comprises the following components: (1) a power-law continuum, (2) an \feii{} pseudo-continuum template developed by \citet[][hereafter T06 template]{Tsuzuki_2006}, (3) two Gaussian profiles to characterize the broad \mgii{} component, (4) a single Gaussian profile to characterize the narrow \mgii{} component, (5) the Balmer continuum and high-order Balmer lines, following \citet{HuChen_2008}. The fitting procedure is executed within specific wavelength windows in the rest frame, spanning 2200--3500 \AA, 3625--3645 \AA, and 4170--4260 \AA. The window 3625--3645 \AA \ is utilized to impose constraints on the Balmer continuum emission due to the absence of pronounced Fe emission here \citep{Wills_1985}. It is difficult to obtain a robust narrow \mgii{} component and a Balmer continuum, but their influence on the final results is negligible. The example of multicomponent spectral fitting in the UV region is illustrated in Figure \ref{fig:fit_example_UV}. 

Throughout our study, we set the boundary between narrow and broad lines at 1200 $\mathrm{km\ s^{-1}}$ \citep{ShenYue_2011}. The \feii{} template was convolved with a Gaussian profile that can be scaled, broadened, and shifted with three free-fitting parameters. We typically use the default initial values for DASpec because the fitting method is insensitive to the first guess. Additionally, we impose certain constraints on the parameter ranges: $-4<\alpha<2$, $100\ \mathrm{km\ s^{-1}}<\mathrm{FWHM_{Fe,conv}}<12000\ \mathrm{km\ s^{-1}}$, $-2500\ \mathrm{km\ s^{-1}}<\mathrm{Shift_{EL}}<2500\ \mathrm{km\ s^{-1}}$, $1200\ \mathrm{km\ s^{-1}}<\mathrm{FWHM_{BEL}}<12000\ \mathrm{km\ s^{-1}}$, $1\ \mathrm{km\ s^{-1}}<\mathrm{FWHM_{NEL}}<1200\ \mathrm{km\ s^{-1}}$, where $\alpha$ is the slope of the power-law continuum, $\mathrm{FWHM_{Fe,conv}}$ represents the FWHM of the convolved Gaussian profile for the \feii{} template, ``EL'' denotes every emission line, ``BEL'' denotes broad emission lines, and ``NEL'' denotes narrow emission lines. We did not consider host galaxy components here. We will discuss the effect of host galaxies in Section \ref{sec:dis_error}. In addition, the choice of the fitting components, the \feii{} templates, and the fitting windows all have influence on the fitting results. We will discuss these issues in Section \ref{sec:dis_error}. We intend to incorporate these effects as additional sources of uncertainty in our analyses. 

\section{Results} \label{sec:results}
In this section, we begin with the process of constructing a quasar sample with \hb{} (hereafter the \hb{} sample). We make the iron correction for \hb-based SMBH masses and perform an exploratory analysis. We then create a quasar sample with \mgii{} (hereafter the \mgii{} sample) and a quasar sample with both \hb{} and \mgii{} (hereafter the \hb{}-\mgii{} sample). Utilizing the \hb{}-\mgii{} sample, we employ the iron-corrected, \hb{}-based SMBH mass $M_{\mathrm{H\beta,corr}}$ to calibrate the iron-uncorrected \mgii{}-based SMBH mass $M_{\text{Mg\,\textsc{ii}}, \mathrm{uncorr}}$, i.e., we derive the iron-corrected $R$--$L$ relation for \mgii{}. Here ``corr'' refers to iron-corrected measurements and ``uncorr'' refers to iron-uncorrected measurements.


\subsection{Iron-corrected, \hb{}-based SMBH Masses} \label{sec:hb_mass}
To explore various single-epoch spectral characteristics associated with \hb{}, we have compiled fundamental parameters from the spectral fitting. These include:
\begin{enumerate}
\item[$\circ$] FWHM and equivalent width (EW) of the broad \hb{} line ($\mathrm{FWHM_{H\beta}}$ and $\mathrm{EW_{H\beta}}$) calculated from the best fit of the double-gaussian model. The DESI spectral resolution is high, so the instrumental broadening effect is negligible.

\item[$\circ$] Flux ratio between \feii{} and \hb{}, denoted as 
\begin{equation} \label{eq:RFe_hb}
R_\mathrm{Fe,\mathrm{H\beta}}=F_\mathrm{Fe}/F_{\mathrm{H\beta}},
\end{equation}
where $F_{\mathrm{H\beta}}$ represents the flux of the broad \hb{} component, and $F_\mathrm{Fe}$ represents the flux of \feii{} spanning from 4434 \r{A} to 4684 \r{A}.

\item[$\circ$] FWHM of \feii{}, which can be expressed as 
\begin{equation} \label{eq:FWHM_fe}
\mathrm{FWHM_{Fe}}=\sqrt{\mathrm{FWHM_{IZw1}}^2+\mathrm{FWHM_{Fe,conv}}^2},
\end{equation}
where the intrinsic FWHM of iron template $\mathrm{FWHM_{IZw1}}$ is constant at 900 $\mathrm{km\ s^{-1}}$ \citep{Vestergaard_2001}. 

\item[$\circ$] EW of \oiii\ $\lambda5007$ ($\mathrm{EW}_{[\text{O\,\textsc{iii}]}}$).

\item[$\circ$] Slope of the power-law continuum ($\alpha$).

\end{enumerate}

\begin{figure}
\epsscale{1.19}
\plotone{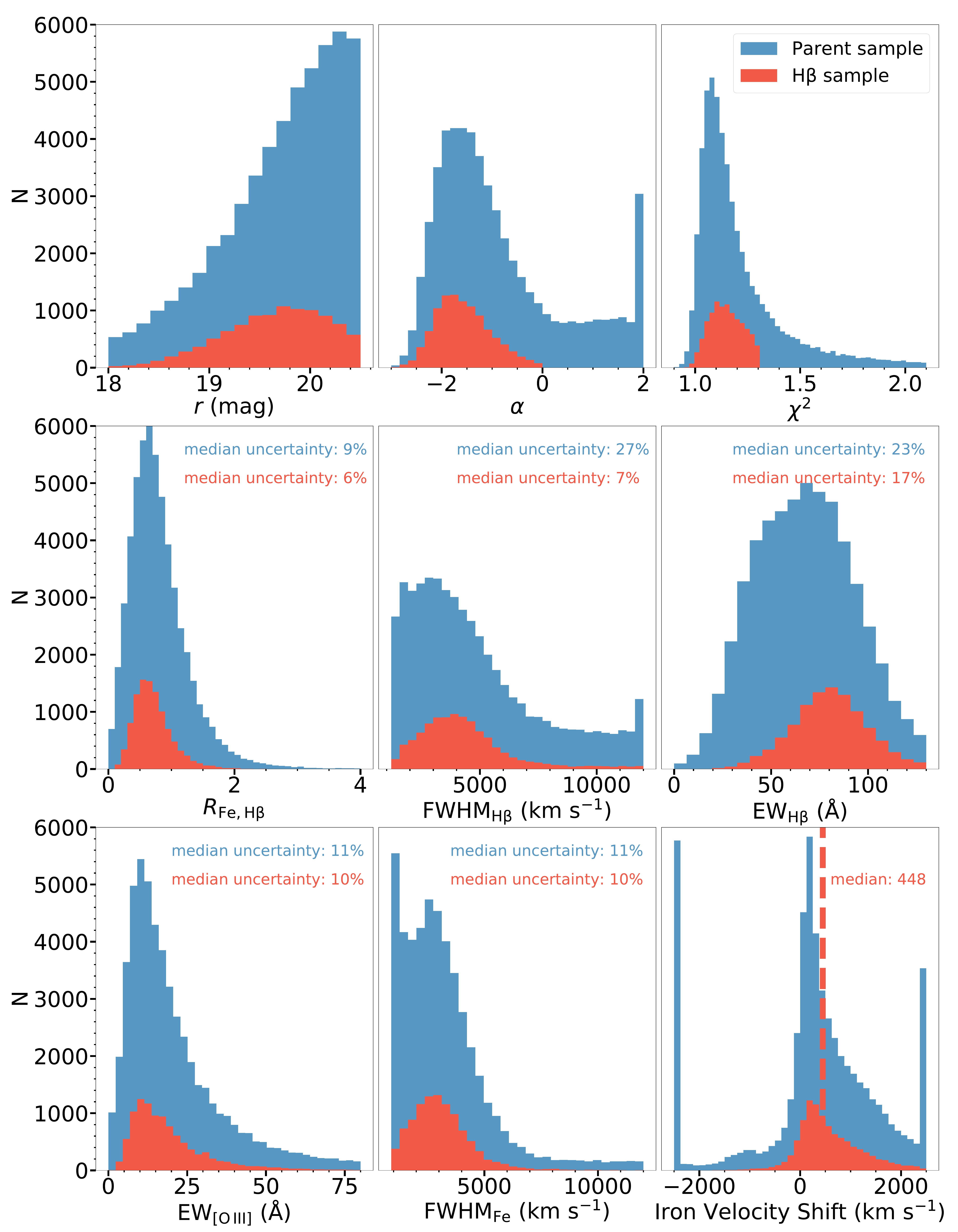}
\caption{Parameter distributions of the parent sample and the \hb{} sample, color-coded with blue and orange. The nine panels display the $r$ magnitude, slope of the power-law continuum, $\chi^2$, flux ratio between \feii{} and \hb{}, FWHM of broad \hb{}, EW of broad \hb{}, EW of \oiii\ $\lambda5007$, FWHM of \feii{}, and velocity shift of \feii{}, respectively. Five panels display the median uncertainty associated with each parameter, indicating that the \hb{} sample has more robust measurements than the parent sample.
\label{fig:Hist_hb}}
\end{figure}

Given the large size of the parent sample, we are able to select quasars with good spectral quality and model fits. Finally, we established a well-defined \hb{} sample with 10,202 quasars using the following criteria:
\begin{equation} \label{eq:hb_criteria}
    \left\{ 
        \begin{array}{ll}
    \mathrm{EW_{H\beta,error}}/\mathrm{EW_{H\beta}}<30\% \\
    R_\mathrm{Fe,\mathrm{H\beta},error}/R_\mathrm{Fe,\mathrm{H\beta}}<30\% \\
    \mathrm{FWHM_{H\beta,error}}/\mathrm{FWHM_{H\beta}}<30\%\\
    \mathrm{EW}_{[\text{O\,\textsc{iii}]}, \mathrm{error}}/\mathrm{EW}_{[\text{O\,\textsc{iii}]}}<30\%\\
    \mathrm{FWHM_{H\beta}}>1500\ \mathrm{km\ s^{-1}} \\
    \chi^2<1.3 \\
    \alpha<0, \\
      \end{array}
   \right.
\end{equation}
where the reduced $\chi^2$ reflects the fitting performance, and ``error'' denotes the 1$\sigma$ uncertainty of the measurements. We simulated 100 spectra for each object, considering the uncertainties associated with all components. The standard deviation of the simulated results for each parameter is treated as the uncertainty.

While most of these selection criteria are geared towards a high S/N, we also impose constraints on $\alpha$ \citep[see Figure 3 in ][]{PanZhiwei_2022} and $\mathrm{FWHM_{H\beta}}$ \citep[see Section 6.2 in][]{Netzer_2013} to focus on blue quasars with typical broad lines, which minimizes contamination from host galaxies. Figure \ref{fig:Hist_hb} illustrates the parameter distributions of the parent sample and \hb{} sample. They are broadly consistent, while the \hb{} sample has more robust measurements. For example, the median relative error of $\mathrm{FWHM_{H\beta}}$ in the \hb{} sample is approximately 7\%, and the median relative error of $L_{5100}$ is around 1\%. The last panel of Figure \ref{fig:Hist_hb} reveals that a significant portion of quasars exhibit redshifted \feii{} emission, with a median velocity of 448 $\mathrm{km\ s^{-1}}$, which is consistent with previous results \citep[e.g.,][]{HuChen_2008}.

In this work, we adopt $\mathrm{log} (R_{\mathrm{H\beta}}/\mathrm{ltd})=1.53+0.51\ \mathrm{log}l_{44}$ as the canonical \hb{}-based $R$-$L$ relation \citep{DuPu_2018}, where ltd is the light day and $l_{44}=L_{5100}/(10^{44}\ \mathrm{erg\ s^{-1}})$. This relation is derived from quasars with low Eddington ratios, and it is almost identical to the relation of \citet{Bentz_2013}.

\begin{figure}
\epsscale{0.97}
\plotone{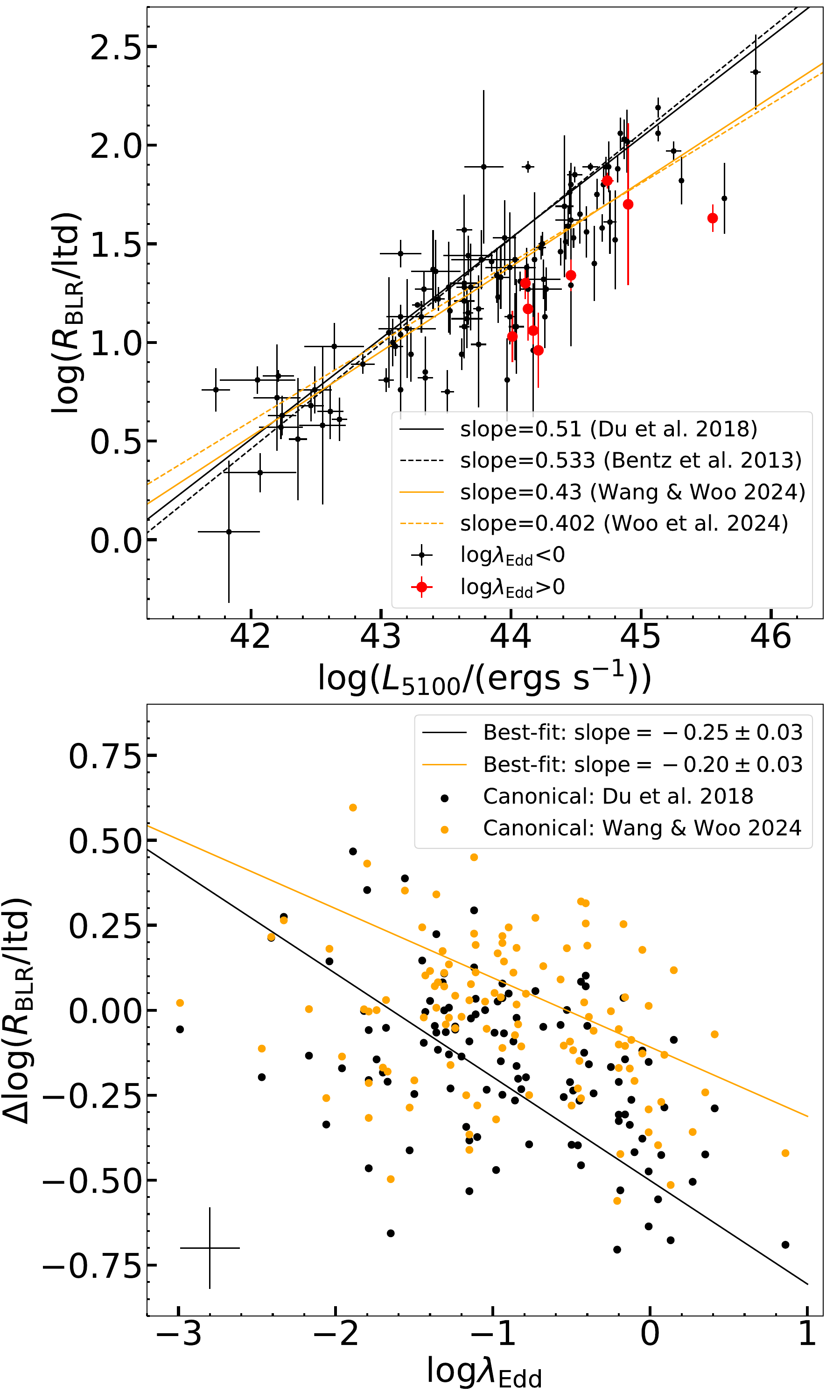}
\caption{Upper: distribution of $R_\mathrm{BLR}$ and $L_{5100}$ from the best average sample in \citet{WangShu_2024}. The black and red dots represent the sub-Eddington and super-Eddington sample, respectively. The lines illustrate four $R$-$L$ relations for comparison. Lower:  correlation between $R_\mathrm{BLR}$ departure and the Eddington ratio. The black and orange colors represent the $R_\mathrm{BLR}$ departure relative to the canonical $R$-$L$ relation from \citet{DuPu_2018} and \citet{WangShu_2024}, respectively. The solid lines indicate the best-fit results from the linear regression. The black cross indicates the typical statistical error bar. Our findings confirm that sub-Eddington and super-Eddington AGNs exhibit systematic offsets in the $R$-$L$ relation, highlighting the necessity for the iron correction.}
\label{fig:different_canonical}
\end{figure}

We also investigate other latest $R$-$L$ relations. For example, \citet{Woo_2024} included more AGNs with moderate to high luminosities and discovered an $R$-$L$ relation with a shallower slope of  $\sim 0.4$. Later, \citet{WangShu_2024} combined this sample with existing samples from the literature to conduct a uniform RM analysis. In the upper panel of Figure \ref{fig:different_canonical}, we present the distribution of $R_\mathrm{BLR}$ and $L_{5100}$ from the best average sample in \citet{WangShu_2024}. The super-Eddington sample, shown in red, is clearly located below the $R$-$L$ relation established by \citet{Bentz_2013} and \citet{DuPu_2018}. It is important to note that the super-Eddington sample is predominantly composed of high-luminosity AGNs, which can explain the shallower slope of $R_\mathrm{BLR}$ versus $L_{5100}$ by \citet{WangShu_2024}. The lower panel of Figure \ref{fig:different_canonical} more clearly illustrates the correlation between the $R_\mathrm{BLR}$ departure and the Eddington ratio. \citet{WangShu_2024} also confirmed that sub-Eddington and super-Eddington AGNs exhibit systematic offsets in the $R$-$L$ relation. These results support that a correction for the Eddington ratio, as indicated by the iron strength, is necessary.

\begin{figure}
\epsscale{1.2}
\plotone{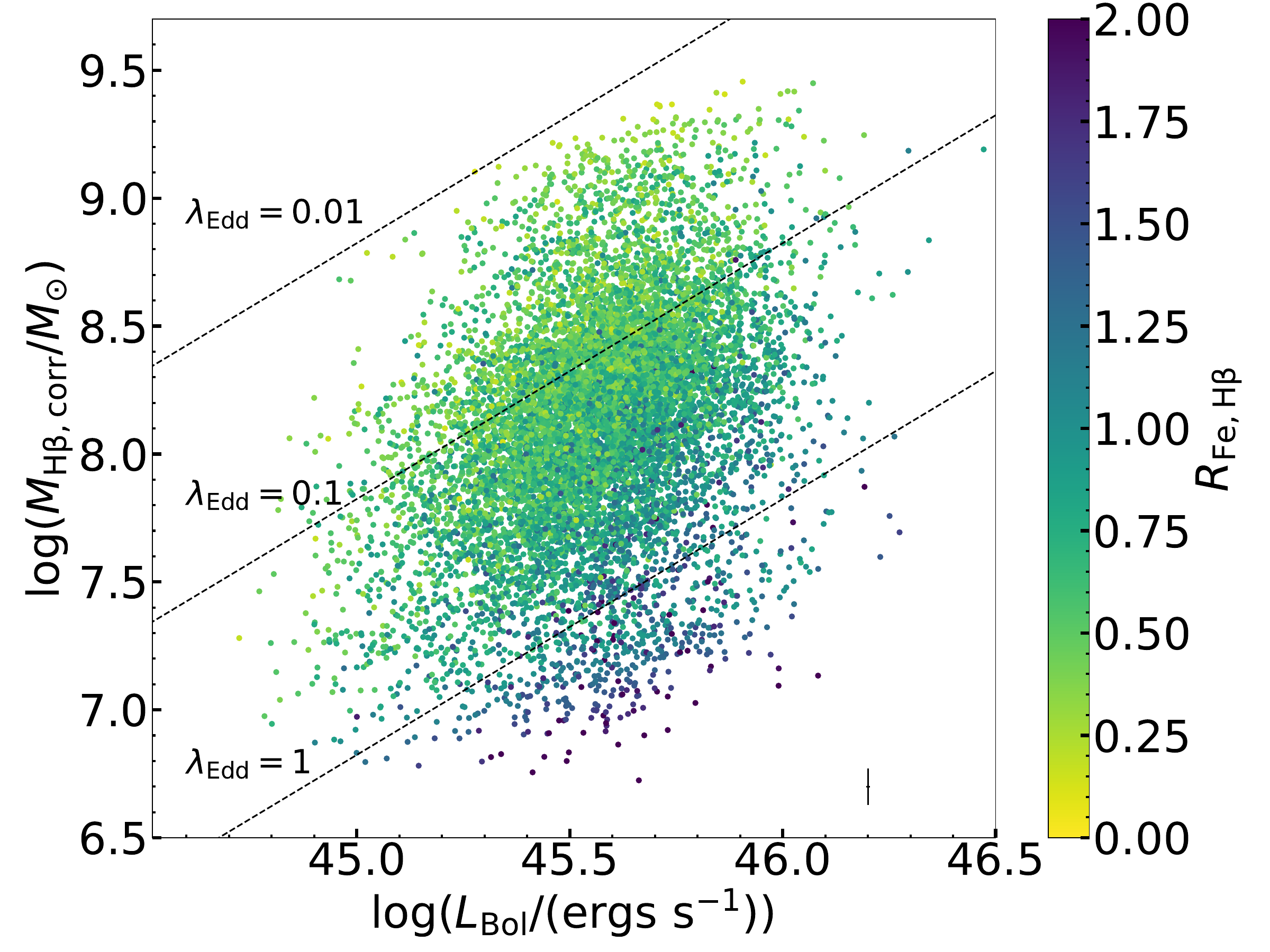}
\caption{The distribution of bolometric luminosity and black hole masses in the \hb{} sample, color-coded by $R_{\mathrm{Fe,\mathrm{H\beta}}}$. The black cross indicates the typical statistical error bar. The dashed lines show the constant Eddington ratios of 0.01, 0.1, and 1, respectively. A reasonable marginal positive correlation is observed between $M_{\mathrm{H\beta}}$ and $L_{\mathrm{Bol}}$. Additionally, at fixed $M_{\mathrm{H\beta}}$, $L_{\mathrm{Bol}}$ increases with larger $R_{\mathrm{Fe,\mathrm{H\beta}}}$, suggesting a relationship between $R_{\mathrm{Fe}}$ and the Eddington ratio.
\label{fig:M_Lbol}}
\end{figure}

We adopt the iron-calibrated $R$-$L$ relation from \citet{DuPu_2019} to estimate the \hb{}-based SMBH mass $M_{\mathrm{H\beta}}$ with the form of
\begin{equation} \label{eq:R-L_Rfe}
\mathrm{log} (R_{\mathrm{H\beta}}/\mathrm{ltd})=1.65+0.45\ \mathrm{log}l_{44}-0.35\ R_{\mathrm{Fe,\mathrm{H\beta}}},
\end{equation}
where the definition of $R_{\mathrm{Fe,\mathrm{H\beta}}}$ is consistent with \citet{DuPu_2019}.
Then Equation \ref{eq:M_RdeltaV} can be employed to derive the iron-uncorrected and iron-corrected, \hb{}-based SMBH masses, $M_{\mathrm{H\beta,uncorr}}$ and $M_{\mathrm{H\beta,corr}}$. We assume that the $f$-factor is 1.1. Its substantial uncertainty will be discussed in Section \ref{sec:dis_error}. The value of 1.1 is converted from the zero point of Equation 1 in \citet{Vestergaard_2009} for consistency. This is also consistent with \citet{Woo_2015}. We will show that the choice of 1.1 provides a fair comparison between \hb{}-based and \mgii{}-based SMBH masses. Then we measure the bolometric luminosity $L_{\mathrm{Bol}}$. Considering that the bolometric correction factor $k_{\mathrm{Bol}}$ is dependent on various parameters such as accretion rate and SMBH mass \citep{JinChichuan_2012}, we adopt Equation 3 from \citet{Netzer_2019}, i.e., $k_{\mathrm{Bol}}=40 \times l_{42}^{-0.2}$, where $l_{42}=L_{5100}/(10^{42}\ \mathrm{erg\ s^{-1}})$. We admit an average uncertainty of $0.2$ dex for the bolometric correction \citep{Netzer_2019}, but this has little impact on the following analysis. Figure \ref{fig:M_Lbol} shows the typical ranges of the bolometric luminosities and SMBH masses in the \hb{} sample. The marginal positive correlation between $M_{\mathrm{H\beta}}$ and $L_{5100}$ is reasonable, as more massive quasars are generally more luminous. Additionally, we compute the Eddington ratio ($L_{\mathrm{Bol}}/L_{\mathrm{Edd}})$, where $L_{\mathrm{Edd}}=1.5 \times 10^{38}(M_{\mathrm{SMBH}}/M_{\odot})$ denotes the Eddington luminosity for gas with solar composition. 

\begin{figure}
\epsscale{1.2}
\plotone{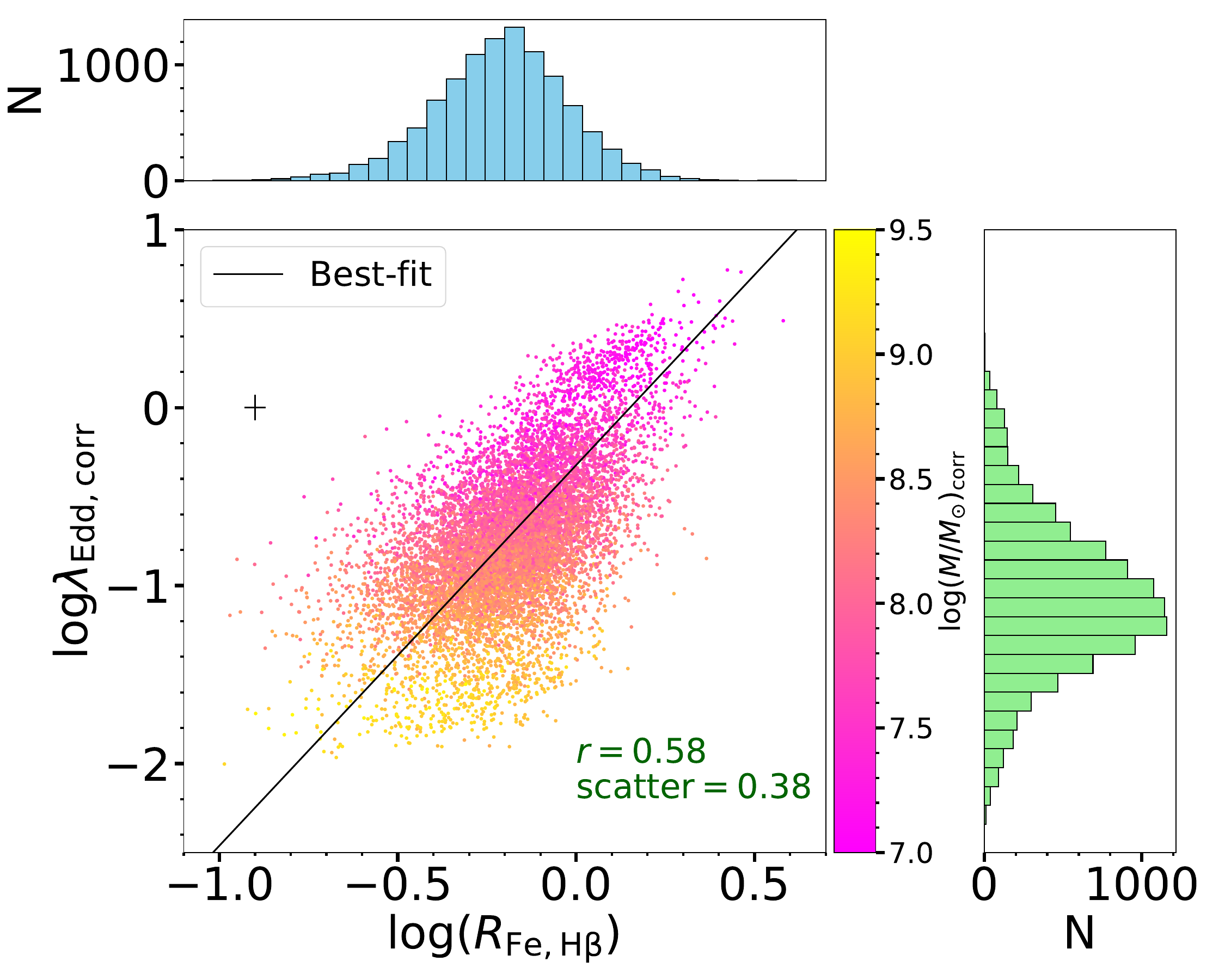}
\caption{Distributions of the Eddington ratio and $R_{\mathrm{Fe}}$, color-coded by $M_{\mathrm{H\beta,corr}}$. The black cross indicates the typical statistical error bar. $r$ is the Pearson correlation coefficient. The black solid line indicates the best-fit result of the linear regression. The rms scatter for the best-fit relationship is about 0.38 dex.} The notable positive correlation in our DESI sample affirms that the $R_\mathrm{Fe,\mathrm{H\beta}}$ term can serve as a reliable tracer for the Eddington ratio.
\label{fig:EddRatio_RFe_hb}
\end{figure}

Figure \ref{fig:EddRatio_RFe_hb} highlights the correlations between the Eddington ratio and $R_\mathrm{Fe,\mathrm{H\beta}}$. In our DESI sample, a notable positive correlation with the rms scatter of 0.38 dex is observed, affirming that the $R_\mathrm{Fe,\mathrm{H\beta}}$ term can serve as a reliable tracer of the Eddington ratio. When we divide the sample into groups with different masses, the marginal correlations persist. It is important to note that the corrected $M_{\mathrm{SMBH}}$ is dependent on $R_\mathrm{Fe,\mathrm{H\beta}}$ according to the formula itself, which enhances the correlation.

\begin{figure*}
\epsscale{1.1}
\plotone{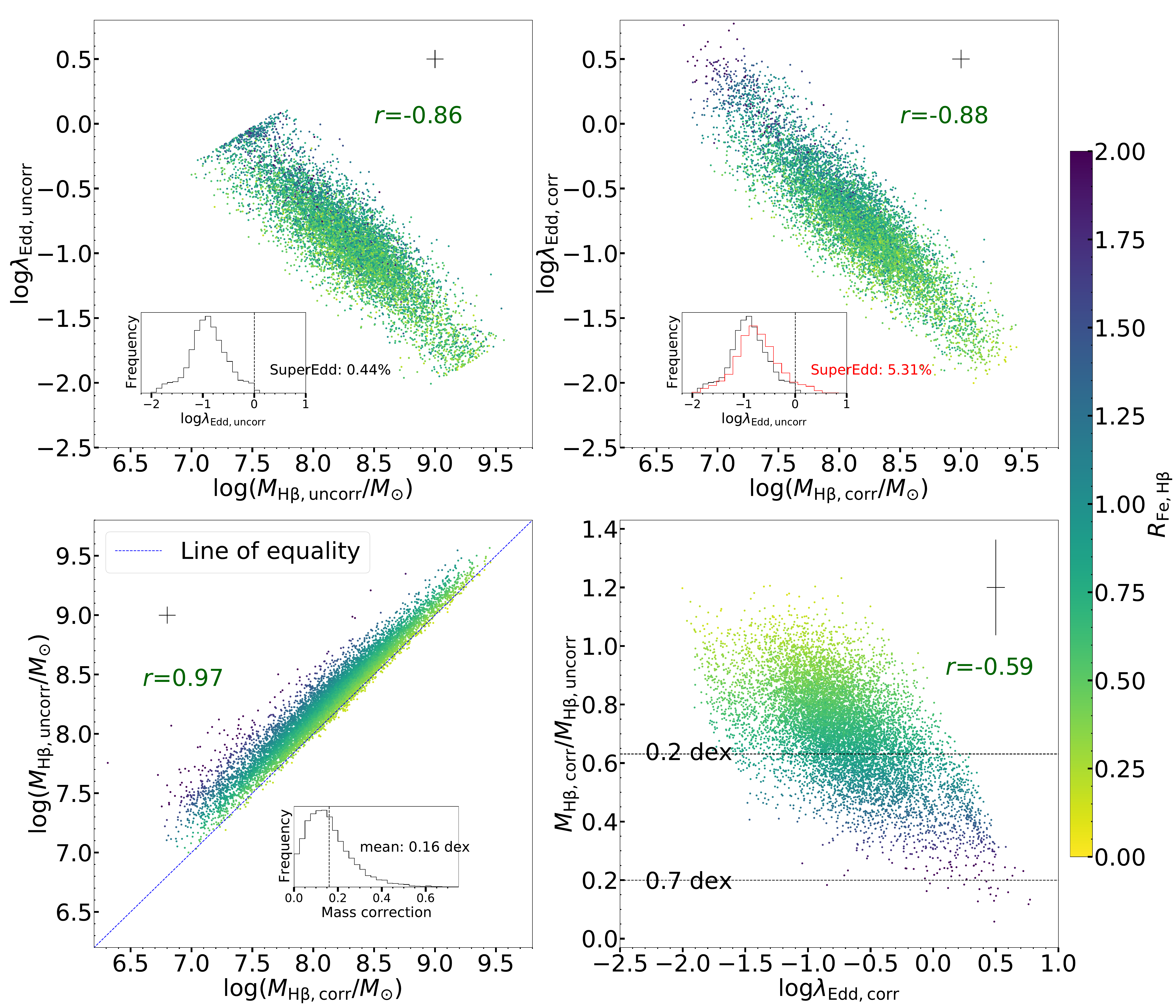}
\caption{Correlations between corrected and uncorrected $M_{\mathrm{H\beta}}$ and Eddington ratio, color-coded by $R_{\mathrm{Fe,\mathrm{H\beta}}}$. The black cross indicates the typical statistical error bar. $r$ is the Pearson correlation coefficient. The blue dashed line shows the line of equality. The upper inset plots show Eddington ratio distributions before and after the correction in black and red, respectively. The dashed vertical line indicated the Eddington ratio of 1. After the correction we estimate that the fraction of super-Eddington quasars is about 5\%, which is significantly higher than the uncorrected value of 0.44\%. The lower panels indicate that the masses of most quasars are overestimated by up to 0.7 dex, depending on the iron strength. The mean correction is about 0.16 dex (equivalent to a factor of 1.5), and the rms and standard deviation are around 0.19 and 0.11 dex, respectively.
\label{fig:4panels}}
\end{figure*}

Correlations between corrected and uncorrected $M_{\mathrm{H\beta}}$ and $L_{\mathrm{Bol}}/L_{\mathrm{Edd}}$ are depicted in Figure \ref{fig:4panels}. The upper panels compare the distributions of the Eddington ratios and SMBH masses before and after the correction. Following the correction, we estimate that the fraction of super-Eddington quasars is around 5\%, significantly higher than the uncorrected value of 0.4\%. The lower panels of Figure \ref{fig:4panels} show that the masses of most quasars are overestimated by up to 0.7 dex, and the mass correction factor shows an anticorrelation with the Eddington ratio. The mean correction is about 0.16 dex (equivalent to a factor of 1.5), and the rms and standard deviation are around 0.19 and 0.11 dex, respectively. Although the mean correction of 0.16 dex is comparable to or smaller than the typical uncertainty of single-epoch SMBH masses, this correction represents a systematic offset driven by the Eddington ratio, and thus is meaningful and necessary.

\begin{figure}
\epsscale{1.15}
\plotone{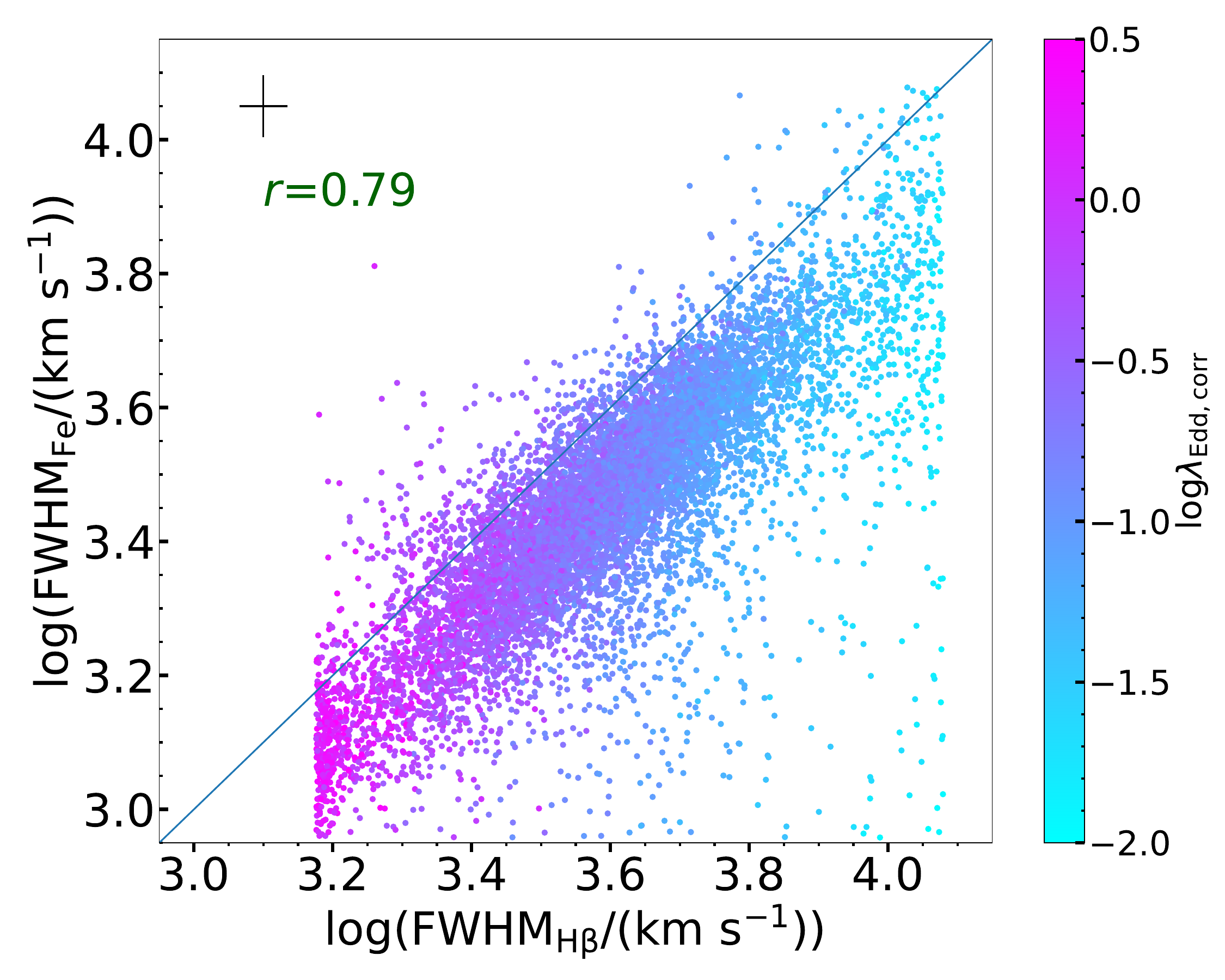}
\caption{The distribution of $\mathrm{FWHM_{Fe}}$ and $\mathrm{FWHM_{H\beta}}$, color-coded by the Eddington ratio. The black cross indicates the typical statistical error bar. $r$ is the Pearson correlation coefficient. The blue dashed line shows the line of equality. This figure confirms that $\mathrm{FWHM_{Fe}}\sim\frac{3}{4}\ \mathrm{FWHM_{H\beta}}$, proposing an association between the optical \feii{} emission and the so-called intermediate-width \hb{} component.
\label{fig:FWHM_Fe_Hb}}
\end{figure}

In Figure \ref{fig:FWHM_Fe_Hb}, the relationships between $\mathrm{FWHM_{Fe}}$, $\mathrm{FWHM_{H\beta}}$, and the Eddington ratio are displayed. The widths of \hb{} are consistently larger than those of \feii{}. This pattern was previously found in the SDSS quasar sample, as discussed in \citet{HuChen_2008}. Their findings indicate that $\mathrm{FWHM_{Fe}}=\frac{3}{4}\ \mathrm{FWHM_{H\beta}}$, proposing an association between the optical \feii{} emission and the so-called intermediate-width \hb{} component that originates from an exterior inflowing region to the BLR region. Furthermore, the relationship between the Eddington ratio and the widths in Figure \ref{fig:FWHM_Fe_Hb} is consistent with previous findings, indicating that highly-accreting AGNs typically exhibit relatively narrow \hb{} lines.

\begin{figure}
\epsscale{1.2}
\plotone{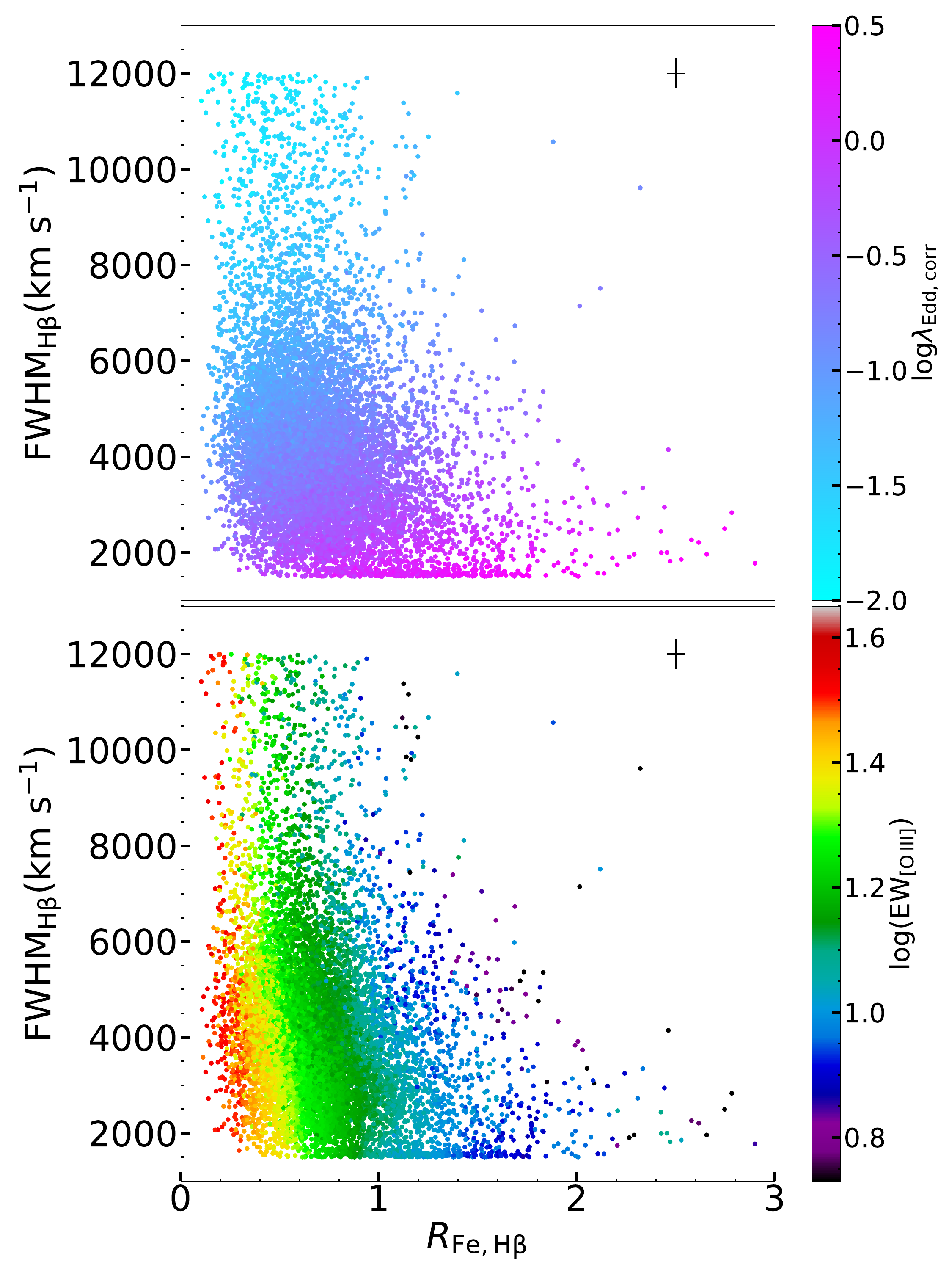}
\caption{Distributions of quasar populations in the EV1 plane. The black cross indicates the typical statistical error bar. In the upper panel, the colors represent the Eddington ratio. In the lower panel, we color-code the points by $\mathrm{EW}_{[\text{O\,\textsc{iii}]}}$, averaged over all nearby objects in a smoothing box of $\Delta R_{\mathrm{Fe}}=0.2$ and $\Delta\mathrm{FWHM_{H\beta}}=1000\ \mathrm{km s^{-1}}$. There is an obvious systematic trend of decreasing $\mathrm{EW}_{[\text{O\,\textsc{iii}]}}$ with increasing $R_{\mathrm{Fe}}$.
\label{fig:EV1}}
\end{figure}

The relation between $R_{\mathrm{Fe}}$ and $\mathrm{FWHM_{H\beta}}$ or $\mathrm{EW}_{[\text{O\,\textsc{iii}]}}$ represents the prominent correlation of Eigenvector 1 (See \citealt{Marziani_2018} for a review), as shown in Figure \ref{fig:EV1}. In the upper panel, coded by the Eddington ratio, there is a marginal anti-correlation between $R_{\mathrm{Fe}}$ and $\mathrm{FWHM_{H\beta}}$. Such a triangle distribution of our DESI sample is consistent with previous results \citep[e.g.,][]{WuQiaoya_2022}. In the low-width regime, the influence of the Eddington ratio on the iron strength becomes more apparent in the horizontal direction. Regarding the vertical direction, the velocity dispersion is commonly attributed to the orientation effect \citep{Marziani_2001,Shen&Ho_2014}. The lower panel is plotted for comparison with Figure 1 from \citet{Shen&Ho_2014}, which used more than 20,000 SDSS quasars within the redshift range of $0.1<z<0.9$ and the luminosity range of $L_{5100}\approx10^{44-45.5}\ \mathrm{erg\ s^{-1}}$. Our results from DESI quasars are well consistent with the SDSS findings, revealing a clear anti-correlation between $\mathrm{EW}_{[\text{O\,\textsc{iii}]}}$ and $R_{\mathrm{Fe}}$.

\subsection{Iron-corrected, \mgii{}-based SMBH Masses} \label{sec:mgii_mass}
As we did for the \hb{} sample, we compile fundamental parameters from the spectral fitting of \mgii. These include:
\begin{enumerate}
\item[$\circ$] FWHM and EW of broad \mgii{} ($\mathrm{FWHM}_{\text{Mg\,\textsc{ii}}}$ and $\mathrm{EW}_{\text{Mg\,\textsc{ii}}}$) calculated from the best fit of the double-gaussian model.
\item[$\circ$] Flux ratio between \feii{} and \mgii{}, denoted as 
\begin{equation} \label{eq:RFe_mgii}
R_{\mathrm{Fe},\text{Mg\,\textsc{ii}}}=F_\mathrm{Fe}/F_{\text{Mg\,\textsc{ii}}},
\end{equation}
where $F_{\text{Mg\,\textsc{ii}}}$ represents the flux of broad \mgii{}, and $F_\mathrm{Fe}$ represents the flux of \feii{} spanning from 2250 \r{A} to 2650 \r{A}.
\item[$\circ$] FWHM of \feii{}.
\item[$\circ$] Slope of the power-law continuum ($\alpha$).
\end{enumerate}

\begin{figure}
\epsscale{1.19}
\plotone{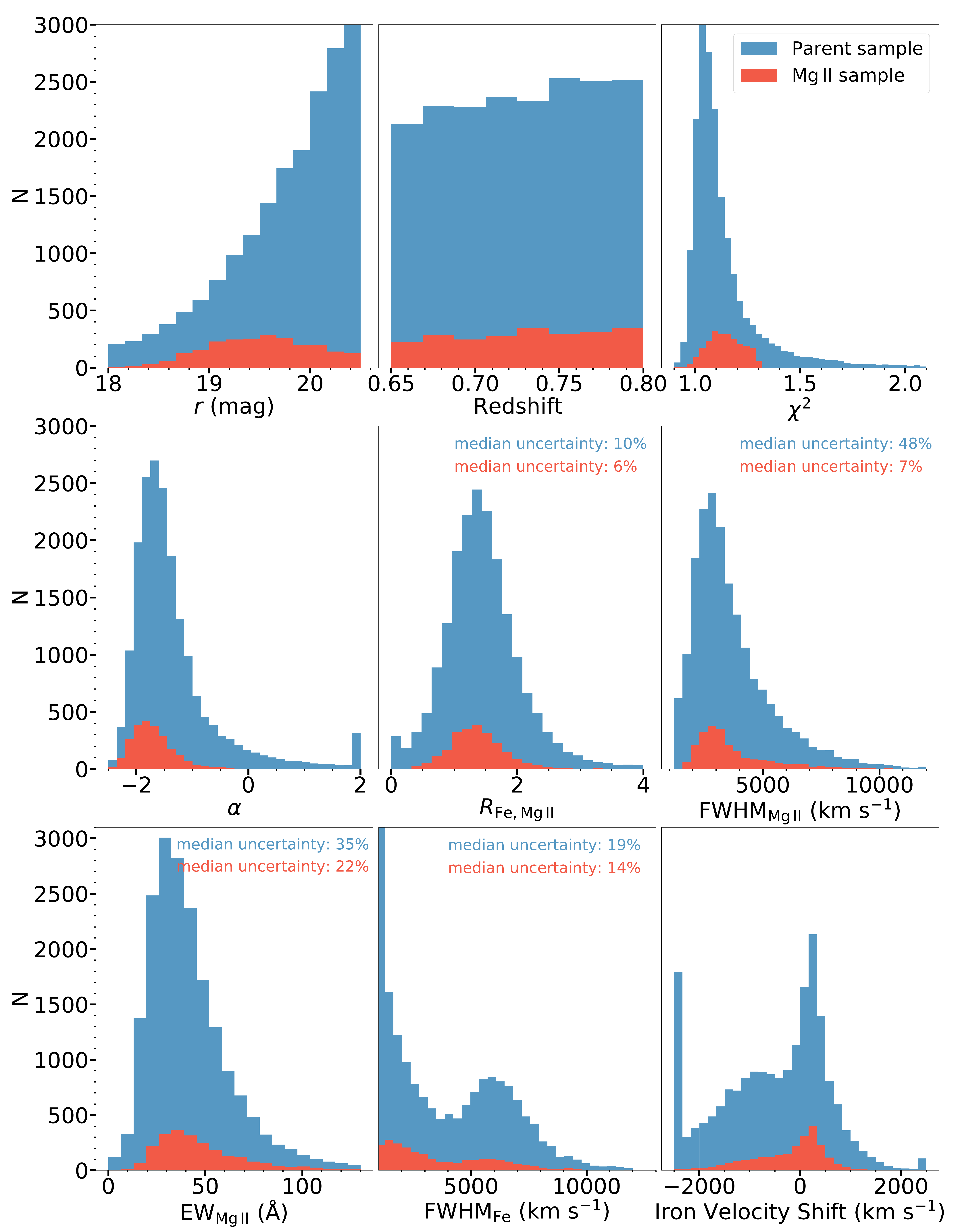}
\caption{Parameter distributions of the parent sample and the \mgii{} sample, color-coded with blue and orange. The nine panels display the $r$ magnitude, redshift, $\chi^2$, slope of the power-law continuum, flux ratio between \feii{} and \mgii{}, FWHM of broad \mgii{}, EW of broad \mgii{}, FWHM of \feii{}, and velocity shift of \feii{}, respectively. Four panels display the median uncertainty associated with each parameter, indicating that the \mgii{} sample has more robust measurements than the parent sample.
\label{fig:Hist_mgii}}
\end{figure}

Following the procedure for the \hb{} sample, we generate a well-defined sample with 2335 quasars (hereafter the \mgii{} sample) using the following criteria:
\begin{equation} \label{eq:mgii_criteria}
    \left\{ 
        \begin{array}{ll}
    0.65<z<0.8 \\
    \mathrm{EW}_{\text{Mg\,\textsc{ii},error}}/\mathrm{EW}_{\text{Mg\,\textsc{ii}}}<30\% \\
    R_\mathrm{Fe,\text{Mg\,\textsc{ii}},error}/R_\mathrm{Fe,\text{Mg\,\textsc{ii}}}<30\% \\
    \mathrm{FWHM}_{\text{Mg\,\textsc{ii},error}}/\mathrm{FWHM}_{\text{Mg\,\textsc{ii}}}<30\%\\
    \mathrm{FWHM}_{\text{Mg\,\textsc{ii}}}>1500\ \mathrm{km\ s^{-1}} \\
    \chi^2<1.3 \\
    \alpha<0. \\
      \end{array}
   \right.
\end{equation}
Figure \ref{fig:Hist_mgii} illustrates the parameter distributions for both the parent sample and the \mgii{} sample. The \mgii{} sample allows more robust measurements than the parent sample. The median relative error of $\mathrm{FWHM_{\text{Mg\,\textsc{ii}}}}$ is about 7\%, and the median relative error of $L_{3000}$ is about 1\%.

Then we perform a crossmatch between the \hb{} sample and the \mgii{} sample, resulting in a \hb{}-\mgii{} sample of 954 quasars. We leverage this \hb{}-\mgii{} sample to explore the relationship between \hb{} and \mgii{}. Specifically, the \hb{}-\mgii{} sample serves as the foundation to calibrate the \mgii{}-based SMBH mass using the estimated iron-corrected, \hb{}-based SMBH mass.

\begin{figure*}
\epsscale{1.11}
\plotone{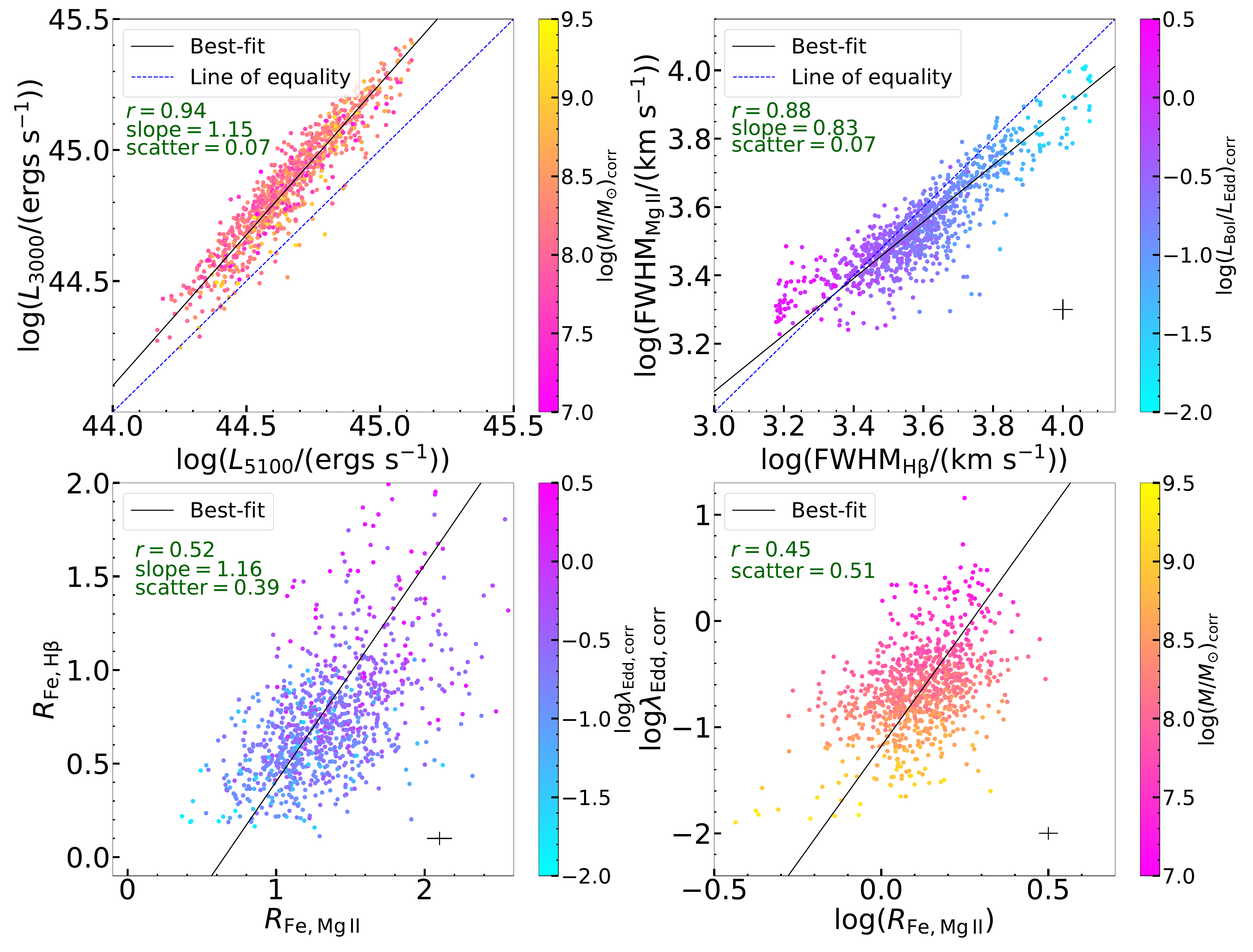}
\caption{Upper left: comparison between $L_{5100}$ and $L_{3000}$. Upper right: comparison between $\mathrm{FWHM_{\text{Mg\,\textsc{ii}}}}$ and $\mathrm{FWHM_{H\beta}}$, color-coded by the Eddington ratio. Lower left: comparison between $R_{\mathrm{Fe,H\beta}}$ and $R_{\mathrm{Fe,\text{Mg\,\textsc{ii}}}}$, color-coded by the Eddington ratio. Lower right: comparison between the Eddington ratio and $R_{\mathrm{Fe,\text{Mg\,\textsc{ii}}}}$, color-coded by $M_{\mathrm{H\beta,corr}}$. The black solid line indicates the best-fit result of the linear regression. The blue dashed line shows the line of equality. The black cross indicates the typical statistical error bar. $r$ is the Pearson correlation coefficient. The slope and rms scatter of the best fit are shown as text in the plot. The slopes of best fit for luminosity, FWHM, and iron strength are all around one, indicating a roughly linear correlation between \mgii{} and \hb{}. The lower panels suggest that both $R_{\mathrm{Fe,H\beta}}$ and $R_{\mathrm{Fe,\text{Mg\,\textsc{ii}}}}$ can trace the Eddington ratio of quasars.} \label{fig:RFe_hb_mgii}
\end{figure*}

We illustrate the comparisons of $L_{5100}$ versus $L_{3000}$, $\mathrm{FWHM_{\text{Mg\,\textsc{ii}}}}$ versus $\mathrm{FWHM_{H\beta}}$, and $R_{\mathrm{Fe,H\beta}}$ versus $R_{\mathrm{Fe,\text{Mg\,\textsc{ii}}}}$ in Figure \ref{fig:RFe_hb_mgii}. In particular, the slope of the best fit for $\mathrm{FWHM_{\text{Mg\,\textsc{ii}}}}$ versus $\mathrm{FWHM_{H\beta}}$ is around 0.83. The sub-linear relation has also been identified in previous studies \citep[e.g.,][]{Woo_2018,Le_2020,Rakshit_2021}. Conversely, some studies have found the roughly linear relation \citep[e.g.,][]{ZuoWenwen_2015}. The results vary depending on different samples, fitting procedures, and templates used. We assume that the slope is roughly one and the relation between \mgii{} and \hb{} is roughly linear. Furthermore, as shown in the lower panels of Figure \ref{fig:RFe_hb_mgii}, both measurements of \feii{} strength trace the Eddington ratio of quasars. The correlation between the Eddington ratio and $R_\mathrm{Fe,\text{Mg\,\textsc{ii}}}$ is only slightly weaker than the correlation shown in Figure \ref{fig:EddRatio_RFe_hb} (The rms scatter is 0.51 dex, compared to 0.38 dex). Therefore, we incorporate $R_\mathrm{Fe,\text{Mg\,\textsc{ii}}}$ to extend the $R$--$L$ relation for \mgii{}.

\begin{figure*}
\epsscale{1.1}
\plotone{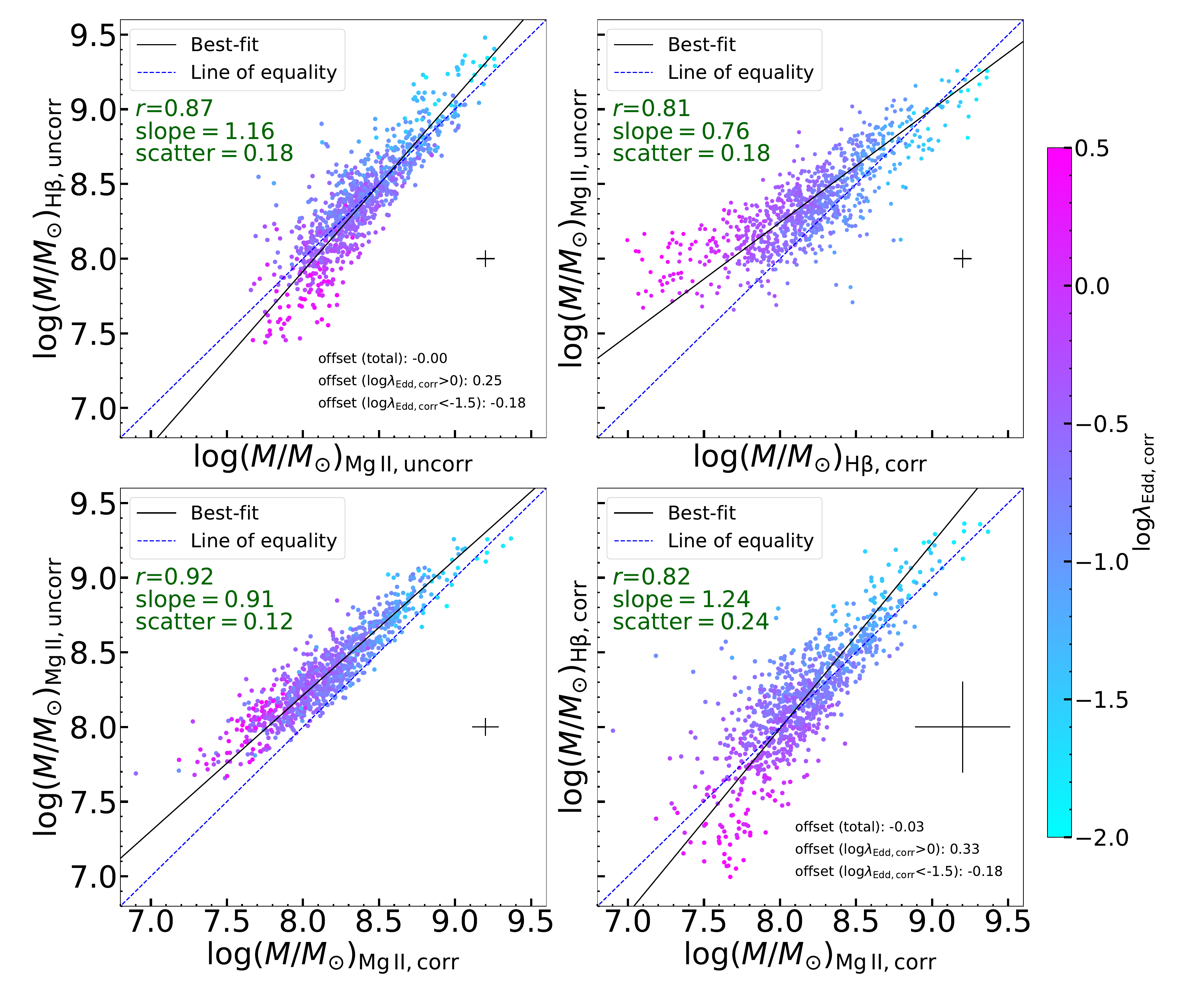}
\caption{Upper left: comparison between the uncorrected $M_{\mathrm{Mg\ II}}$ and uncorrected $M_{\mathrm{H\beta}}$. Upper right: comparison between uncorrected $M_{\mathrm{Mg\ II}}$ and corrected $M_{\mathrm{H\beta}}$. Lower left: comparison between the uncorrected $M_{\mathrm{Mg\ II}}$ and corrected $M_{\mathrm{Mg\ II}}$. Lower right: comparison between the uncorrected $M_{\mathrm{Mg\ II}}$ and corrected $M_{\mathrm{Mg\ II}}$. The black solid line indicates the best-fit result of the linear regression, and the grey shaded region indicates its 1$\sigma$ uncertainty. The black crosses in the first three panels indicate the typical statistical error bar. In the lower right panel, we show the typical total uncertainty, including the statistical error and an additional 0.3 dex uncertainty from the f factor. $r$ is the Pearson correlation coefficient. The blue dashed line shows the line of equality. The iron correction effect on $M_{\mathrm{Mg\ II}}$ is similar to that on $M_{\mathrm{H\beta}}$. For the entire sample, the average correction is about a factor of 1.5, while for the super-Eddington sample, it is approximately a factor of 2.3.}\label{fig:Mhb_Mmgii}
\end{figure*}

The uncorrected virial mass based on \mgii{} $M_{\mathrm{\text{Mg\,\textsc{ii}},uncorr}}$ is determined using the following equation \citep[][hereafter VO09]{Vestergaard_2009}:
\begin{equation} \label{eq:m_mgii}
M=10^{6.86}\left(\frac{\mathrm{FWHM_{\text{Mg\,\textsc{ii}}}}}{10^3\ \mathrm{km\ s^{-1}}}\right)^2\left(\frac{L_{3000}}{10^{44}\ \mathrm{erg\ s^{-1}}}\right)^{0.5}\ M_{\odot},
\end{equation}
where $L_{3000}$ is the monochromatic luminosity ($\lambda L_{\lambda}$) at 3000 \r{A}. The upper left panel of Figure \ref{fig:Mhb_Mmgii} shows a comparison between the $M_{\mathrm{\text{Mg\,\textsc{ii}},uncorr}}$ and $M_{\mathrm{H\beta, uncorr}}$. These two tracers are consistent with each other. The mean offset is less than 0.003 dex, which proves that the choice of the viral factor $f$ of 1.1 in Equation \ref{eq:M_RdeltaV} is appropriate. In the upper right panel of Figure \ref{fig:Mhb_Mmgii}, especially in the high Eddington ratio regime, it is noticeable that $M_{\mathrm{\text{Mg\,\textsc{ii}},uncorr}}$ tends to be systematically higher than $M_{\mathrm{H\beta,corr}}$ due to the previously mentioned iron correction effect for $M_{\mathrm{H\beta}}$.

We utilize $M_{\mathrm{H\beta,corr}}$ as a reference and perform a multivariable linear regression with the following form:
\begin{equation} \label{eq:M_mgii_corrrected}
\begin{split}
\mathrm{log}\frac{M_{\mathrm{\text{Mg\,\textsc{ii}},corr}}}{M_{\odot}}=a+b\times\mathrm{log\frac{FWHM_{\text{Mg\,\textsc{ii}}}}{km\ s^{-1}}} \\
+c\times\mathrm{log}\frac{L_{3000}}{10^{44}\ \mathrm{erg\ s^{-1}}}+d\times R_\mathrm{Fe,\text{Mg\,\textsc{ii}}},
\end{split}
\end{equation}
where we fix $b=2$. This is to compare with \hb{}-based masses for consistency. This choice has been recommended and used by many previous studies \citep[e.g.,][]{Vestergaard_2009,ShenYue_2011,Woo_2018,Le_2020,YuZhefu_2023}. The best-fit results are: $a=1.14\pm0.03$, $c=0.46\pm0.03$, $d=-0.34\pm0.02$, which are very similar to the parameters of Equation \ref{eq:R-L_Rfe}. This similarity suggests a potential relationship between optical iron and UV iron emission. We show the comparison between $M_{\mathrm{\text{Mg\,\textsc{ii}},uncorr}}$ and $M_{\mathrm{\text{Mg\,\textsc{ii}},corr}}$ in the lower left panel of Figure \ref{fig:Mhb_Mmgii}. The iron correction effect on $M_{\mathrm{\text{Mg\,\textsc{ii}}}}$ is similar to that on $M_{\mathrm{H\beta}}$. For the entire sample, the average correction is about a factor of 1.5, while for the super-Eddington sample, it is approximately a factor of 2.3. Finally, the lower right panel of Figure \ref{fig:Mhb_Mmgii} demonstrates that $M_{\mathrm{\text{Mg\,\textsc{ii}},corr}}$ and $M_{\mathrm{H\beta, corr}}$ show a good overall consistency. The average offset between these two estimators is 0.03 dex, with a scatter of 0.24 dex. We observe a mean offset of 0.3 dex in the lowest mass range and 0.2 dex in the highest mass range. This discrepancy arises from the sub-linear relationship between \( \mathrm{FWHM_{\text{Mg\,\textsc{ii}}}} \) and \( \mathrm{FWHM_{H\beta}} \). Specifically, \( \mathrm{FWHM_{\text{Mg\,\textsc{ii}}}} \) tends to be slightly larger than \( \mathrm{FWHM_{H\beta}} \) when the lines are very narrow, while it is typically larger than \( \mathrm{FWHM_{H\beta}} \) when the lines are broad. \citet{Le_2020} discussed this effect and suggested an additional correction term based on the profile of \mgii{}. In our analysis, we subtracted the narrow component of \mgii{}, which is a different approach from that used by \citet{Le_2020}. This approach makes the above discrepancy more prominent. Nonetheless, the two iron-corrected mass estimators are generally consistent within the typical total uncertainty of 0.31 dex. We again emphasize the importance of calibrating \( M_{\mathrm{\text{Mg\,\textsc{ii}}}} \) based on the accretion state, as indicated by the iron strength mentioned above.

\subsection{SMBH Mass Estimation of DESI Quasars at $0.6<z<1.6$} \label{sec:DESI_mass}

With the iron-corrected single-epoch SMBH mass estimator for \mgii{}, we measure the SMBH masses of DESI quasars at $z<2$. Our plan is to integrate the \hb{} and \mgii{} estimators to provide a reliable BH mass estimation in the future. We will collaborate with other DESI SMBH mass studies and release the results as part of the DESI DR1 value-added catalog (VAC). In this paper, we restrict the SMBH mass measurements to the redshift range of $0.6 < z < 1.6$, only utilizing the derived \mgii{} estimator.

\begin{deluxetable*}{cccc}    
\tablecaption{Fits Catalog Format} 
\tablewidth{0pt}
\tablehead{ Column  & Format & Units & Description}
\startdata    \label{tab1:mass}
RA & float64 & degree & Right Ascension  \\
DEC & float64 & degree & Declination  \\
TARGETID & int64 & & Target ID \\
PROGRAM & string & & Program ID \\
HEALPIX & int64 & & HEALPix ID \\
REDSHIFT & float64 & & Redshift from FSF \\
RMAG & float64 & mag & $22.5-2.5\log_{10}(\text{FLUX\_R})$ \\
L3000\_DAS & float64 & $1 \times 10^{44}$ erg/s & Luminosity at 3000Å (DAS) \\
L3000\_DAS\_ERR & float64 & $1 \times 10^{44}$ erg/s & Error of L3000\_DAS \\
EW\_DAS & float64 & \AA & Equivalent Width of Mg II Emission Line (DAS) \\
EW\_DAS\_ERR & float64 & \AA & Error of EW\_DAS \\
FWHM\_DAS & float64 & km/s & Full Width at Half Maximum of Mg II Emission Line (DAS) \\
FWHM\_DAS\_ERR & float64 & km/s & Error of FWHM\_DAS \\
RFE\_DAS & float64 & & Strength of Iron Emission in UV (DAS) \\
RFE\_DAS\_ERR & float64 & & Error of RFE\_DAS \\
LOGMASS\_DAS\_Pan25 & float64 & $\mathrm{M}_{\odot}$ & Iron-corrected SMBH Mass based on Pan25 (DAS) \\
LOGMASS\_DAS\_Pan25\_ERR & float64 & ... & Error of LOGMASS\_DAS\_Pan25 \\
LOGMASS\_DAS\_VO09 & float64 & ... & SMBH Mass based on VO09 (DAS) \\
LOGMASS\_DAS\_VO09\_ERR & float64 & ... & Error of LOGMASS\_DAS\_VO09 \\
LOGMASS\_DAS\_Shen11 & float64 & ... & SMBH Mass based on Shen11 (DAS) \\
LOGMASS\_DAS\_Shen11\_ERR & float64 & ... & Error of LOGMASS\_DAS\_Shen11 \\
LOGMASS\_DAS\_Le20 & float64 & ... & SMBH Mass based on Le20 (DAS) \\
LOGMASS\_DAS\_Le20\_ERR & float64 & ... & Error of LOGMASS\_DAS\_Le20 \\
LOGMASS\_DAS\_Yu23 & float64 & ... & SMBH Mass based on Yu23 (DAS) \\
LOGMASS\_DAS\_Yu23\_ERR & float64 & ... & Error of LOGMASS\_DAS\_Yu23 \\
L3000\_FSF & float64 & $1 \times 10^{44}$ erg/s & Luminosity at 3000Å (FSF) \\
FWHM\_FSF & float64 & km/s & Full Width at Half Maximum of MgII Emission Line (FSF) \\
LOGMASS\_FSF\_VO09 & float64 & $\mathrm{M}_{\odot}$ & SMBH Mass based on VO09 (FSF) \\
\enddata
\tablecomments{This table is available in its entirety in the machine-readable format. ``FSF'' indicates the spectral measurements obtained from FastSpecFit, while ``DAS'' signifies the results derived from DASpec in this paper.}
\end{deluxetable*}

Since DESI has not released an official quasar catalog, we first build a quasar catalog as follows. We choose a primary sample from an Iron VAC (v2.0) generated by FastSpecFit (J. Moustakas et al. in preparation)\footnote{\url{https://fastspecfit.readthedocs.io/en/latest/iron.html}}. FastSpecFit is a stellar continuum and emission-line modeling code designed for DESI. This VAC provides continuum and emission line information crucial for SMBH mass estimation. This dataset will be publicly released as part of DESI DR1. We select the sample from the DESI main survey and apply constraints such as $0.6<z<1.6$, $r<22$, spectype==``qso'', and survey== ``main''. There would be misclassified objects with large uncertainty in the catalog. By adding constraints to the uncertainties, users could remove them. In the future we will update our SMBH mass catalog after the release of the official DESI quasar catalog.

We apply the same fitting procedure mentioned above and utilize Equation \ref{eq:M_mgii_corrrected} to estimate $M_{\mathrm{\text{Mg\,\textsc{ii}},corr}}$ with proper error propagation. Finally, we analyze 490,648 quasars listed in Table \ref{tab1:mass}. Among them, approximately 35\% quasars have a mass error smaller than 0.5 dex. Additionally, we estimate $M_{\mathrm{\text{Mg\,\textsc{ii}}}}$ using the fitting results from the FastSpecFit VAC and Equation \ref{eq:m_mgii}. In order to provide users with the convenience of choosing their preferred estimation methods, we also provide masses based on the formula derived by \citet[][hereafter Shen11]{ShenYue_2011}, \citet[][hereafter Le20]{Le_2020}, and \citet[][hereafter Yu23]{YuZhefu_2023}. We will discuss the details and make the comparisons between different SMBH mass estimators in Section \ref{sec:dis_compare}.


\section{Discussion}   \label{sec:discuss}
\subsection{Uncertainties of the SMBH Masses} \label{sec:dis_error}
The SMBH mass measurements derived above depend on the fitting procedure. Hence, a key question is to test the robustness of the result. We discuss the uncertainties in this subsection.

Spectral fitting windows can influence the fitting results. We have chosen a broad window spanning 3750 to 5500 \AA\ for the optical region to include enough features. Compared to the windows used in \citet{DuPu_2019}, we have added a window of 3750--4000 \AA\ to better decompose the continuum and \feii{}. Therefore, the choice of optical window is not expected to introduce a significant uncertainty. However, in the UV region, the fitting windows has some impact on the results. This sensitivity arises because part of the \feii{} line lies beneath the \mgii{} line, making the relative flux ratio more sensitive. Another issue is related to the UV \feii{} itself. While this study uses a single velocity shift parameter to regulate the iron template (consistent with most studies), individual UV iron lines may have different shifts \citep{Vestergaard_2001}. Consequently, different fitting windows may be dominated by different iron lines, which introduces additional uncertainties. Furthermore, for most quasars in the \hb{}-\mgii{} sample, the \mgii{} line is located at the edge of spectra that have poorer S/N ratios on the blue side of \mgii{}. This effect leads to a weak constraint on the continuum. To estimate the additional uncertainty in a simple way, we randomly select 100 quasars from the \hb{}-\mgii{} sample. For each quasar, we fit three times with three setups of UV fitting windows, the 2200--3500 \AA, 2200--3400 \AA, and 2200--3600 \AA\ windows. We do not test changing the left limit because it has already almost reached the edge of spectral coverage. For simplicity, we do not change the fitting windows of 3625--3645 \AA\ and 4170--4260 \AA. We acknowledge that what we estimate is a lower limit of the additional uncertainty. Finally, we consider the standard deviation of 100 objects as the uncertainty. Overall, we estimate that $\mathrm{FWHM}_{\text{Mg\,\textsc{ii}}}$, $L_{3000}$, and $R_{\mathrm{Fe,\text{Mg\,\textsc{ii}}}}$ have additional uncertainties of 5\%, 2\%, and 6\%, respectively.

The \mgii{} feature is sensitive to the \feii{} emission that extends beneath the emission line. Previous studies have highlighted that the choice of the UV \feii{} template can introduce significant biases to the results \citep[e.g.,][]{Kurk_2007,Woo_2018,Shin_2019,YuZhefu_2021}. One frequently used UV iron template is from \citet[][referred to as the VW01 template]{Vestergaard_2001}. The T06 template has also been widely used \citep[e.g.,][]{Sameshima_2017,Woo_2018,Shin_2019}. The main differences lie in their behaviors around the \mgii{} line \citep[see Figure 1 in][]{Lai_2024_XQz5}. The T06 template incorporates the \feii{} contribution beneath the \mgii{} line, employing a model spectrum based on a photoionization calculation of the BLR clouds. In contrast, the VW01 template lacks \feii{} flux in this region, leading to an overestimation of the actual \mgii{} flux. However, determining the optimal template remains challenging. \citet{Lai_2024_XQz5} considered four empirical and semi-empirical UV iron templates: VW01 template, T06 template, the template from \citet{Bruhweiler_2008}, and the template from \citet{MejiaRestrepo2016}. They derived the final line properties and measurement uncertainties from the mean results obtained using these four different iron templates. Following the same methodology, we fit the 100 quasars with four setups of the iron templates mentioned above. We estimate that there is a systematic difference of about 29\% between the mean $R_{\mathrm{Fe,\text{Mg\,\textsc{ii}}}}$ based on four templates. It is risky to define such a systematic difference as an uncertainty because we do not know the true value. Since the mean $R_{\mathrm{Fe,\text{Mg\,\textsc{ii}}}}$ based on the T06 template has an intermediate result, we adopt the T06 template for this work. Besides, the influence of templates on $\mathrm{FWHM}_{\text{Mg\,\textsc{ii}}}$ and $L_{300}$ is about 4\%. Therefore, we acknowledge that the lower limit of the additional uncertainty caused by iron templates is about 4\%.

The choice of fitting components for emission lines can slightly affect the results. Our study utilized a line model comprising one narrow Gaussian component and two broad Gaussian components for \hb{}. In contrast, some studies use one narrow and three broad Gaussian components to characterize \hb{} \citep[e.g.,][]{Rakshit_2020,YangJinyi_2023}. Given our primary focus on flux ratios, the impact of the number of \hb{} components is minimal. Regarding \oiii{}, previous works sometimes use one Gaussian for the narrow component (the ``core'' component) and a second Gaussian profile for the potentially broadened and blueshifted component (the ``wing'' component). The velocity shift and dispersion of the narrow \hb{} are tied to those of the \oiii{} ``core'' component. However, it is challenging to distinguish the two components robustly. More important, for most objects, the ``core'' component dominates the whole \oiii{}. Therefore, we use one Gaussian to characterize \oiii{}. Similarly, we choose two, not three, broad Gaussian components for \mgii{}. We have also explored the necessity of including Balmer continuum emission. We observed a significant degeneracy between the Balmer and power-law continuum when the fitting windows of the UV region are narrow. Considering this, we adopt the current broad fitting windows to better constrain the Balmer continuum. In summary, our choice of emission line components is reasonable and is expected to introduce negligible uncertainties.

We also test the influence of the host galaxy. Based on a spectrophotometric decomposition technique introduced by Sun et al. (in preparation), we randomly fit 1000 quasars at $z\sim0.6$ from Table \ref{tab1:mass}. The results suggest that the host galaxy fraction anticorrelates with the quasar luminosity. In the range of $10^{45}\ \mathrm{erg\ s^{-1}}<L_{\mathrm{Bol}}<10^{46}\ \mathrm{erg\ s^{-1}}$, the average optical host fraction is about 20\% and the UV host fraction is about 5\%, which is consistent with recent results \citep[e.g.][]{ShenYue_2011,Jalan_2023}. In addition, \citet{Jalan_2023} found anticorrelations between the host galaxy fraction and iron strength, Eddington ratio, and redshift. Therefore, for the majority of objects in our \hb{}-\mgii{} sample at $0.65<z<0.8$, host galaxy contamination should have little effect on the final results.

In order to incorporate the above effects, we add an additional uncertainty of 10\% for $\mathrm{FWHM}$ and $R_{\mathrm{Fe}}$. The typical final uncertainties of the two parameters are 0.055 dex and 0.052 dex, respectively. We apply a correction of 20\% for $L_{5100}$ to exclude the host galaxy contribution. This introduces an additional uncertainty 25\% and leads to a typical final uncertainty of 0.109 dex. Similarly, we apply a correction of 5\% to $L_{3000}$, and its typical final uncertainty is 0.044 dex. Following the same procedure as introduced in Section \ref{sec:mgii_mass}, we obtain the best fit of $d=-0.29\pm0.02$ when we fix the parameter $c=0.46$. The value of $-0.29$ is roughly consistent with $-0.34$ within a 2 $\sigma$ level, confirming the need for the iron correction.  

In the context of a specified line width, the virial f-factor is contingent on the geometry and kinematics of the BLR gas, exhibiting significant variations across different objects. The virial f-factor can range from approximately $0.5$ \citep{Graham_2011} to $1.1$ \citep{Woo_2013}. Its determination often involves comparing the virialized SMBH masses to the expected SMBH masses derived from the local $M$--$\sigma$ relation. Note that different works use different velocity indicators, such as $\sigma$ or $\mathrm{FWHM}$. Additionally, the $M$--$\sigma$ relation is dependent on galaxy bulge properties, suggesting that different virial coefficients should be applied based on the bulge classification of the host galaxy, with values around $1.1$ for classical bulges and $0.6$ for pseudobulges \citep{Ho_2014}. However, choosing the most appropriate f-factor based on different types is challenging, leading to intrinsic scatter in the f-factor. Consequently, a mean value is often employed for SMBH mass estimations.  Our study assumed a mean f-factor of $1.1$, converted from \citet{Vestergaard_2009}. We acknowledge an intrinsic scatter of approximately $0.3$ dex \citep{ShenYue_2024}, constituting a major source of uncertainty in single-epoch SMBH mass estimations. It is essential to note that, since our focus is on the systematic iron correction of the \mgii{}-based SMBH mass, the absolute f-factor of \hb{} estimators only affects the absolute mass measurements but not the calibration formula. 

Altogether, we estimate that the overall systematic uncertainty of a single-epoch SMBH mass is approximately 0.4-0.5 dex, significantly larger than the typical measurement error. The primary purpose of this work is to reduce the systematic offset rather than the scatter of mass estimations by introducing the iron correction. The derived iron-corrected $R$-$L$ relation for \mgii{} is robust.

\subsection{Comparison with Previous \mgii{}-based SMBH Mass Estimators} \label{sec:dis_compare}
The differences of SMBH mass estimators can be attributed to numerous factors, such as sample construction, fitting procedures, the f-factor, the form of the formula, and more. In this subsection, we delve into the complexities by determining masses using various estimators and drawing comparisons to investigate the systematic differences.

\begin{figure}
\epsscale{1.18}
\plotone{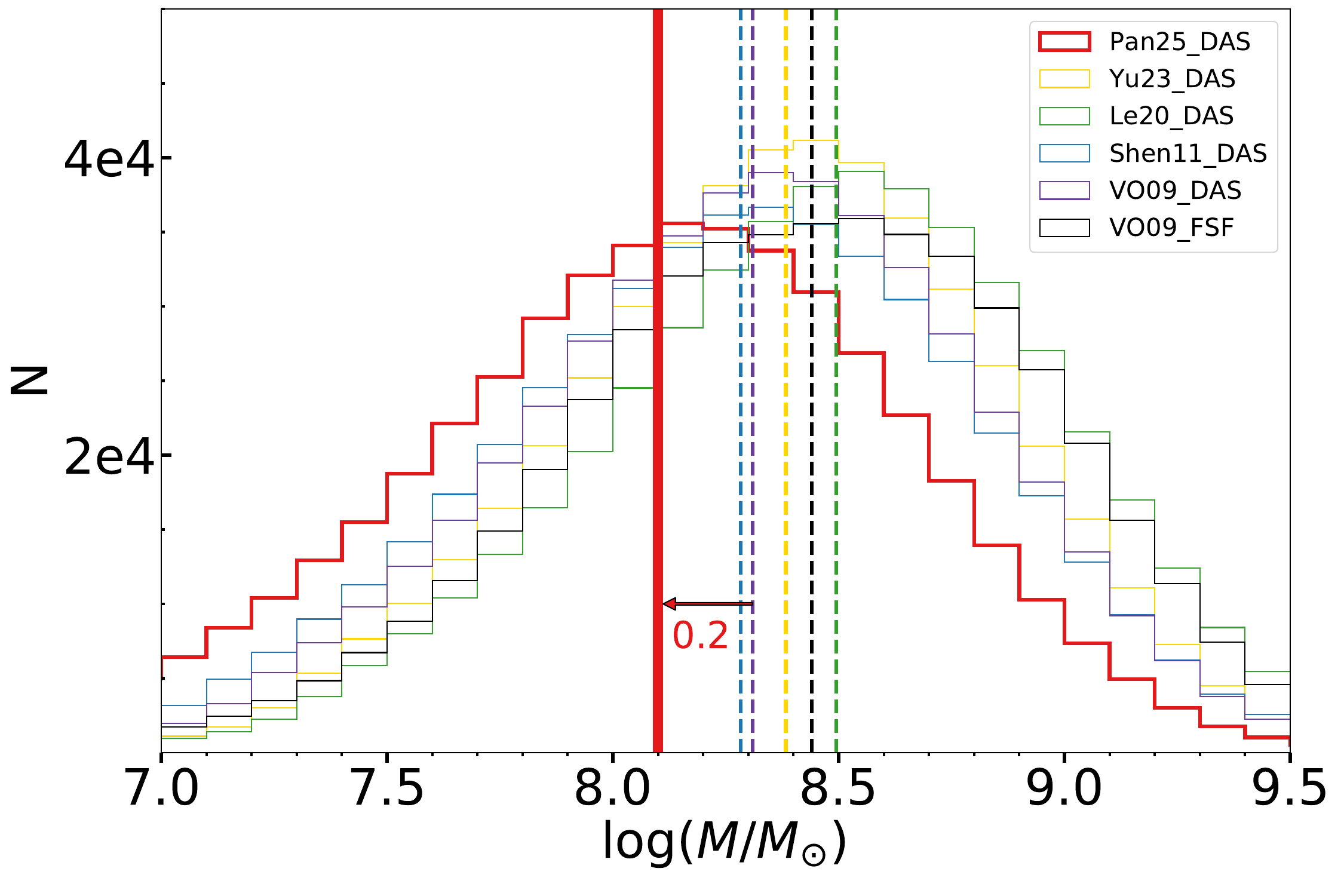}
\caption{\mgii{}-based SMBH mass distributions of different estimators. The red solid line represents the median mass estimated by Equation \ref{eq:M_mgii_corrrected} in this paper (Pan25), which is the only iron-corrected estimator among them. The black dashed line represents the median mass estimated by VO09 with the fitting results from the FastSpecFit. The remaining colors denote the estimators VO09, Shen11, Le20, and Yu23 with the fitting results derived by this paper. The average mass reduction caused by the iron correction is about 0.2 dex.}
\label{fig:compare_M}
\end{figure}

As listed in Table \ref{tab1:mass}, we have selected four estimators from the literature for comparison. These estimators have parameters as follows.
\begin{equation}
    \begin{aligned}
        \mathrm{VO09:}\ (a,b,c,d) &= (0.86, 2, 0.5, 0) \\
        \mathrm{Shen11:}\ (a,b,c,d) &= (0.74, 2, 0.62, 0) \\
        \mathrm{Le20:}\ (a,b,c,d) &= (1.04, 2, 0.5, 0) \\
        \mathrm{Yu23:}\ (a,b,c,d) &= (1.01, 2, 0.39, 0) \\
        \mathrm{Pan25:}\ (a,b,c,d) &= (1.14, 2, 0.46, -0.34). \\
    \end{aligned}
\end{equation}
We estimate the \mgii{}-based masses based on these formulas and display the mass distributions and median masses of different estimators in Figure \ref{fig:compare_M}. VO09 (purple) serves as a reference estimator, while Pan25 (red) indicates the measurements using Equation \ref{eq:M_mgii_corrrected} in this work. Overall, the average iron correction is about 0.2 dex.

Shen11 (blue) allowed the luminosity index $c$ to vary, yet they employed the same calibration as VO09, resulting in a negligible systematic offset between the outcomes of VO09 and Shen11. Le20 (green) utilized a sample with a large luminosity range to calibrate the best-fit parameters of the \mgii{} estimator. Their best-fit model fixed $b=2$ and $c=0.5$ with a large fitted value of $a$, which is consistent with the results reported by \citet{Woo_2018}. Thus, the median mass of Le20 is systematically larger by around 0.2 dex than that of Shen11, which has been reported and discussed in \citet{Le_2020}. Shen11 subtracted the narrow component of \mgii{} and derived systematically higher $\mathrm{FWHM}_{\text{Mg\,\textsc{ii}}}$, which led to a smaller calibration result of $a$ for a given fiducial mass compared to Le20. Yu23 (yellow) used their latest RM results and derived $c=0.39$, which is shallower than the canonical value of 0.5. This estimator obtains slightly larger median BH mass than VO09 due to the different calibration and best-fit parameters. The black dashed line is the result from VO09 with FastSpecFit fitting results, which did not consider the \feii{} template. This led to an overestimation of the continuum luminosity and $\mathrm{FWHM}_{\text{Mg\,\textsc{ii}}}$, and thus an overestimation of SMBH masses.

Due to the systematic discrepancies caused by various factors (e.g., iron correction, virial factor f, components of \mgii{}, fixed or free parameters, and \feii{} templates), caution should be taken when selecting SMBH mass estimators. Therefore, we provide the original FWHM and luminosity values in Table \ref{tab1:mass}, and users can easily use their preferred mass estimators. In the future, more and more RM and interferometric results will help to better study the effect of the iron correction and to derive a better estimator.

\subsection{Implications} \label{sec:dis_impli}
We further investigate whether the iron correction is also valid at higher redshifts, particularly in the range of $z>6$ when the universe is less than one billion years old. Recent observations of quasars at high redshift are difficult to reconcile with black hole formation and growth scenarios because they require extremely massive seeds and rapidly accrete gas at the Eddington or even super-Eddington rate (see \citealt{Inayoshi_2020} for a review). For example, SDSS J010013.02+280225.8 is an extremely massive quasar at $z=6.3$ with an estimated SMBH mass of $1.2\times10^{10}\ M_{\odot}$ \citep{WuXuebing_2015}. \citet{WangFeige_2018} reported a quasar at $z=7.0$ with a mass of approximately $1.3\times10^{9}\ M_{\odot}$. The James Webb Space Telescope (JWST) also observed a quasar at $z=7.1$ with a mass of about $1.5\times10^{9}\ M_{\odot}$ \citep{Bosman_2023} and an AGN at $z=8.5$ with a mass of about $1.5\times10^{8}\ M_{\odot}$ \citep{Kokorev_2023}. Besides, many JWST programs including ASPIRE \citep[e.g.,][]{YangJinyi_2023,WangFeige_2024}, CEERS \citep[e.g.,][]{Finkelstein_2023,Larson_2023}, and UNCOVER \citep[e.g.,][]{Bezanson_2022,Goulding_2023}, have revealed a large population of AGNs in the early universe. The ongoing JWST observations will continue to identify a larger population of massive quasars, suggesting different scenarios for the assembly of SMBHs.

Most of these high-redshift luminous quasars host billion solar-mass black holes and have high Eddington ratios \citep[e.g.,][]{YangJinyi_2023,Larson_2023,Goulding_2023}, it is natural to ask whether this population requires the iron correction mentioned above. If the iron-corrected $R$-$L$ relations for \hb{} and \mgii{} are confirmed in the early universe, it suggests that current measurements of high-redshift SMBH masses are significantly overestimated by around 0.4 dex on average. To some extent, the tension of the required massive seeds could be relieved by such a mass reduction. Therefore, we plan to collect a high-redshift sample and study its properties to address this question in the future.

\section{Summary} \label{sec:sum}
In this study, we assembled a parent sample of more than 55,000 type 1 quasars in the redshift range of $0.25 < z < 0.8$ from a DESI internal data release. We used versatile spectral fitting techniques to measure the spectral properties of the quasars, including continuum and line emission in the UV and optical ranges. We then estimated SMBH masses of the quasars and corrected the mass measurements using the iron emission line strength. 

Our best \hb{} sample consists of more than 10,000 quasars. We calculated their SMBH masses using the iron-corrected $R$--$L$ relation that accounts for the accretion state, which caused a mass reduction of up to 0.7 dex. Our analysis revealed that approximately 5\% of super-Eddington quasars exist in this sample, showcasing a notable positive correlation and affirming the reliability of the $R_\mathrm{Fe,\mathrm{H\beta}}$ term as a tracer for the Eddington ratio. We also confirmed a relation between the $\mathrm{FWHM_{Fe}}$ and $\mathrm{FWHM_{H\beta}}$ emissions that suggests a connection between the optical \feii{} emission and the intermediate-width \hb{} component. Our findings agree closely with the previous SDSS results, highlighting a clear anti-correlation between $\mathrm{EW}_{[\text{O\,\textsc{iii}]}}$ and $R_{\mathrm{Fe}}$.

Furthermore, we constructed a \hb{}-\mgii{} sample comprising around 1000 type 1 quasars at $0.65 < z < 0.8$. Using the iron-corrected, \hb{}-based SMBH mass, we calibrated \mgii{}-based SMBH masses and established the iron-corrected $R$--$L$ relation for \mgii{}. The derived relation takes the form $\mathrm{log} (R_{\text{Mg\,\textsc{ii}}}/\mathrm{ltd})\sim0.46\ \mathrm{log}l_{44}-0.34\ R_{\mathrm{Fe}}$. Notably, their iron-corrected SMBH masses are approximately 0 to 0.4 dex lower than classic single-epoch SMBH masses. If this iron correction still holds in the early universe, the tension of very massive SMBH seeds could be alleviated. We also conducted comparisons between our results and those of previous estimators. In general, they are consistent, but some minor systematic differences still exist. Leveraging the derived \mgii{} estimator, we provided SMBH masses for around 0.5 million DESI quasars at $0.6<z<1.6$, with a plan to expand this dataset in the future. 

\section*{Data availability} 
All the material needed to reproduce the figures of this publication is available at this site: \url{https://doi.org/10.5281/zenodo.14673134}.

\section*{acknowledgments}   \label{sec:ack}
We acknowledge support from the National Key R\&D Program of China (2021YFA1600404), the National Science Foundation of China (12225301), and the China Manned Space Project (CMS-CSST-2021-A05, CMS-CSST-2021-A06). We thank DESI Internal Reviewers (Mar Mezcua and Jinyi Yang) for very helpful discussions and comments and thank DESI PubBoard Handler (Vanina Ruhlmann-Kleider) for timely help. We thank Jian-Min Wang, Pu Du, Luis C. Ho, Yuxuan Pang, and Masafusa Onoue for very helpful discussions and comments.  HZ acknowledge the supports from National Key R\&D Program of China (grant Nos. 2023YFA1607800, 2022YFA1602902, 2023YFA1608100) and the National Natural Science Foundation of China (NSFC; grant Nos. 12120101003, 12373010, 12233008) and the Strategic Priority Research Program of the Chinese Academy of Science (Grant no. XDB0550100).

This material is based upon work supported by the U.S. Department of Energy (DOE), Office of Science, Office of High-Energy Physics, under Contract No. DE–AC02–05CH11231, and by the National Energy Research Scientific Computing Center, a DOE Office of Science User Facility under the same contract. Additional support for DESI was provided by the U.S. National Science Foundation (NSF), Division of Astronomical Sciences under Contract No. AST-0950945 to the NSF’s National Optical-Infrared Astronomy Research Laboratory; the Science and Technology Facilities Council of the United Kingdom; the Gordon and Betty Moore Foundation; the Heising-Simons Foundation; the French Alternative Energies and Atomic Energy Commission (CEA); the National Council of Humanities, Science and Technology of Mexico (CONAHCYT); the Ministry of Science and Innovation of Spain (MICINN), and by the DESI Member Institutions: https://www.desi.lbl.gov/collaborating-institutions. 
Any opinions, findings, and conclusions or recommendations expressed in this material are those of the author(s) and do not necessarily reflect the views of the U. S. National Science Foundation, the U. S. Department of Energy, or any of the listed funding agencies.
The authors are honored to be permitted to conduct scientific research on Iolkam Du’ag (Kitt Peak), a mountain with particular significance to the Tohono O’odham Nation.

\facilities{DESI}
\software{Astropy\citep{Astropy_2013,Astropy_2018,Astropy_2022}, Topcat\citep{Taylor_2005.TOPCAT}, FastSpecFit\citep{fastspecfit23}, DASpec}

\newpage
\bibliography{DESI_BH}{}

\bibliographystyle{aasjournal631}

\end{document}